\documentclass[11pt,a4paper,oldfontcommands,openany]{memoir}
\usepackage[english]{babel}
\usepackage[a4paper,outer=3cm,top=2cm,bottom=2cm]{geometry}

\usepackage[compsci,hyperref]{atlas-preamble}

\usepackage{mathrsfs}

\usepackage{csquotes} %
 	
\usepackage{url}
\usepackage{pgfplots}
\usepackage{caption}
\usepackage{subcaption}
\graphicspath{{./figures/}{./figures/gswd/}}
\DeclareUnicodeCharacter{2212}{−}
\usepgfplotslibrary{groupplots,dateplot}
\usetikzlibrary{patterns,shapes.arrows}
\pgfplotsset{compat=newest}
\usepackage{tikz-cd}

\usepackage[
backend=bibtex,
style=alphabetic,
sorting=nyt,
giveninits=true, %
isbn=false, %
]{biblatex} %
\AtEveryBibitem{%
  \clearfield{eprint}%
}

\usepackage{amsthm}

\usepackage{thmtools}
\usepackage{thm-restate}

\theoremstyle{plain}
\newtheorem{theorem}{Theorem} %
\newtheorem{lemma}[theorem]{Lemma} %
\newtheorem{corollary}[theorem]{Corollary} %
\newtheorem{assumption}{Assumption} %

\theoremstyle{definition}
\newtheorem{definition}{Definition} %

\theoremstyle{remark}
\newtheorem*{remark}{Remark} %

\renewcommand{\vec}[1]{\bm{#1}} %
\newcommand{\mat}[1]{\mathbf{#1}} %

\newcommand{\tauht}{\hat \tau_{ht}} %
\newcommand{\taunet}{\hat \tau_{net}} %

\usepackage{titlesec}

\usepackage{cleveref} %

\makeatletter
\if@twoside
  \copypagestyle{chapter}{Ruled}
\else
  \copypagestyle{chapter}{ruled}
\fi
\makeoddhead{chapter}{}{}{}
\makeevenhead{chapter}{}{}{}
\makeheadrule{chapter}{\textwidth}{0pt}
\copypagestyle{abstract}{empty}
\makechapterstyle{bianchimod}{%
  \chapterstyle{default}
  \renewcommand*{\chapnamefont}{\normalfont\Large\sffamily}
  
  \renewcommand*{\printchaptername}{%
    \chapnamefont\centering\@chapapp}

    }

\chapterstyle{bianchimod} %

\titleformat{\paragraph}%
  [runin]%
  {\normalfont\bfseries}%
  {\theparagraph}%
  {1em}%
  {}%
  []%

\titleformat{\subsection}%
  [hang]%
  {\sffamily\large\bfseries}%
  {\thesubsection}%
  {1em}%
  {}%
  []%
  
 \titleformat{\section}%
  [hang]%
  {\sffamily\Large\bfseries}%
  {\thesection}%
  {1em}%
  {}%
  []%

\pretitle{\vspace{0pt plus 0.7fill}\begin{center}\HUGE\sffamily\bfseries}
\posttitle{\end{center}\par}
\preauthor{\par\begin{center}\let\and\\\Large\sffamily}
\postauthor{\end{center}}
\predate{\par\begin{center}\Large\sffamily}
\postdate{\end{center}}

\def\@advisors{}
\newcommand{\advisors}[1]{\def\@advisors{#1}}
\def\@department{}
\newcommand{\department}[1]{\def\@department{#1}}
\def\@thesistype{}
\newcommand{\thesistype}[1]{\def\@thesistype{#1}}

\renewcommand{\maketitlehookb}{\vspace{1in}%
  \par\begin{center}\Large\sffamily\@thesistype\end{center}}
\renewcommand{\maketitlehookd}{%
  \vfill\par
  \begin{flushright}
    \sffamily
    \@advisors\par
    \@department, ETH Z\"urich
  \end{flushright}
}

\title{Randomized Controlled Trials Under Influence:\\ Covariate Factors and  Graph-Based Network Interference}
\author{Tassilo Schwarz\thanks{[lastname]@maths.ox.ac.uk}}
\advisors{Advisors: Prof.\ Dr.\ Rasmus Kyng, Federico Soldà}
\department{Department of Computer Science}

\date{November 2021}

\begin{document}

\maketitle
\newpage

\begin{abstract}
Randomized controlled trials are not only the golden standard in medicine and vaccine trials but have spread to many other disciplines like behavioral economics, making it an important interdisciplinary tool for scientists. 

When designing randomized controlled trials, how to assign participants to treatments becomes a key issue. In particular in the presence of covariate factors, the assignment can significantly influence statistical properties and thereby the quality of the trial. Another key issue is the widely popular assumption among experimenters that participants do not influence each other -- which is far from reality in a field study and can, if unaccounted for, deteriorate the quality of the trial.

We address both issues in our work. After introducing randomized controlled trials bridging terms from different disciplines, we first address the issue of participant-treatment assignment in the presence of known covariate factors. Thereby, we review a recent assignment algorithm that achieves good worst-case variance bounds.

Second, we address social spillover effects. Therefore, we build a comprehensive graph-based model of influence between participants, for which we design our own average treatment effect estimator $\hat \tau_{net}$. We discuss its bias and variance and reduce the problem of variance minimization to a certain instance of minimizing the norm of a matrix-vector product, which has been considered in literature before. Further, we discuss the role of disconnected components in the model's underlying graph.
\end{abstract}

\setlength{\parindent}{0pt}
\setlength{\parskip}{0.3em}
\newpage
\section*{Acknowledgements}
I am grateful to my advisors Prof.\ Rasmus Kyng and Federico Soldà for their support throughout this work, in particular their feedback on new ideas and mathematical advice.

Working on an interdisciplinary topic required support and inspiration from many sides. I am grateful to Prof.\ Stefan Bechtold who introduced me to randomized controlled trials and with whom I have been working on that topic for the last years. I appreciate the insightful discussion on randomized controlled trials in behavioral economics with Dr.\ Tobias Gesche.

Working on an interdisciplinary topic where little theoretic work exists (RCTs with social spillover effects) requires making many fundamental decisions. I am grateful to my advisors for providing me with the necessary guidelines and recommendations. 

\section*{Notes on the public version}
After giving an introduction to randomized controlled trials, this work is two-fold: On the one hand, it reviews recents progress for optimal experiment designs (in particular the Gram-Schmidt Walk Design, \Cref{sec:gswd-main}). On the other hand, it addresses the issue of interference among units by introducing a novel model and estimator for the average treatment effect (\Cref{sec:network-effects}).

For readers only interested the latter, we recommend to still read the spectral discussion (\Cref{sec:spectral-discussion}), as this provides a good intuition for the formal arguments made in \Cref{sec:network-effects}.

This work reflects the research undertaken during my bachelor thesis project ``Randomized Controlled Trials: Review and Network Effects'' in the Computer Science Department at ETH Zürich.

\newpage
{\hypersetup{hidelinks}
\tableofcontents
}
\newpage

\chapter{Introduction}

\section*{Origin and Relevance}

Randomized controlled trials (RCTs) play a key role in assessing the effectiveness of human-related treatments. While the first trials comparing two treatments dates back to 600 B.C. \cite{grimes-ClinicalResearchAncient-1995a}, Fisher formalized randomized experiments in a mathematical sense in his seminal work in 1935 \cite{fisher-DesignExperiments-1935}. The first modern \emph{randomized} controlled clinical trial \cite{StreptomycinTreatmentPulmonary-1948} was published 13 years later. The ``inventor'' of the modern randomized controlled trial, Sir Austin Bradford Hill, described the following three advantages of using randomness: (1) it eliminates any personal bias in the construction of treatment groups, (2) it eliminates overcompensation, that might be introduced if aware of personal bias and (3) it allows to refer to randomness to argue that, with high probability, the groups are unbiased \cite{hill-ClinicalTrial-1952}.

\section*{Key idea}
The idea behind RCTs is very simple: First, a quantity of interest is defined that can be measured for each participant (in the following called \emph{unit}), and that is hypothesized to be influenced by the treatment. Such a quantity of interest could be the infection rate for a certain disease.\\
 Then, units get randomly partitioned into a treatment group, $Z^+$, and a control group, $Z^-$. Units in $Z^+$ get a certain treatment, like a vaccine or medicine, while units in $Z^-$ get a placebo. Finally, the quantity of interest is compared between these two groups -- allowing to reason about the effectiveness of the treatment. 
 
 This intuitive explanation shall be enough for the arguments presented in the introduction, as a formal definition will be given in \Cref{sec:rct-formal-def}.
 
 \section*{Challenge}
 The key question is how to perform the random allocation of units. While the easiest and widely used possibility is to assign each unit independently with probability $p = \frac{1}{2}$ into treatment / control, Hill already realized a key challenge in his early paper: What if there are multiple subgroups that differ from each other? Intuition tells us to balance the share of treatment / control units within each of the subgroups, for the simple reason: If we have a significant imbalance and the membership of a group influences the quantity of interest (we say that the membership to a group is \emph{predictive of the outcome}), we might end up measuring that imbalance of the groups instead of the treatment's effect. We will dive more into this idea of balance within subgroups throughout \Cref{sec:RCT-def-and-review}.
  
\section*{Early assumptions}  \label{sec:SUTVA}
Rubin coined an assumption made by Cox in 1958 as the so-called \emph{stable unit treatment value assumption (SUTVA)}: Any unit's observed outcome depends only on the treatment it was assigned to, not on any other unit. 

\begin{displayquote}
	``If unit $i$ is exposed to treatment $j$, the observed value of $Y$ will be $Y_{ij}$; that is, there is no interference between units (Cox 1958, p. 19) leading to different outcomes depending on the treatments other units received [...]'' -- \cite{rubin-RandomizationAnalysisExperimental-1980}
\end{displayquote}
This assumption remains dominant in many RCTs to date, even though it might be inaccurate in many instances. We will address this issue below and provide more accurate models beyond SUTVA in \Cref{sec:network-effects}.

 \section*{RCTs Today}
 
 RCTs have spread to many other disciplines to perform either field studies (i.e., studies in the real world) or laboratory experiments, where the environment can be controlled. These disciplines include psychology, behavioral economics, educational research, and agricultural science. In medicine, they remain to date the ``golden standard for ascertaining the efficacy and safety of a treatment'' \cite{kabisch-RandomizedControlledTrials-2011}.
 
 Surprisingly, the in 1958 established stable unit treatment value assumption (SUTVA) remains to date a dominant assumption in many RCTs. Especially in field studies, where the experimenter has limited influence on the surroundings, the assumption that units do not affect each other can be inaccurate and lead to wrong conclusions.

Another important property that can affect the quality of an RCT is the balancedness of \textdef{covariate factors}, i.e., properties of units that might affect their treatment's outcome. Large trials often rely on the law of large numbers, hoping that and i.i.d.\ design\footnote{a design in which each unit gets allocated with probability $\frac{1}{2}$ to either treatment or control group, independent of the others} avoids imbalances of predictive characteristics between the treatment and control group.
One such large RCT is the medical trial performed to assess the performance of the BNT162b2 mRNA Covid-19 Vaccine \cite{polack-SafetyEfficacyBNT162b2-2020}: With over 35.000 participants, most subgroups have approximately an equal share in the treatment and control group. However, the subgroup \emph{Native Hawaiian or other Pacific Islander} is not balanced at all:
\begin{center}
 \begin{tabular}{l | c c } 

 Subgroup & \# in treatment & \# in control \\ [0.5ex] 
 \hline
 \specialcell{Native Hawaiian or other \\ Pacific Islander (n=76)} & 50 & 26\\ 

\end{tabular}
\end{center}   

If this subgroup were predictive of the treatment outcome, such an imbalance could distort the outcome. In this case, this is only a small subgroup, so that it has a negligible impact on the overall average outcome -- but might have a significant impact for members of this subgroup.

 \section*{Overview}
 The goal of this work is to both discuss the idea of balance within subgroups as well as to give a framework to make RCTs work beyond the stable unit treatment value assumption (SUTVA).
 
In \Cref{sec:RCT-def-and-review}, we will introduce RCTs formally and dive into the idea of balance within subgroups. We examine this idea both from the perspective of medical RCT literature as well as from the perspective of a recent paper by Harshaw et al.\ \cite{harshaw-BalancingCovariatesRandomized-2021}. This paper provides an efficient algorithm for finding a design with two desirable properties: unbiasedness and low variance of a specific treatment effect estimator.
 
\Cref{sec:network-effects} covers RCTs beyond SUTVA. We motivate the need for frameworks capable of dealing with RCTs in the presence of peer influence, review literature on this issue, define a new model and estimator, and analyze its error. Under basic assumptions, we can give a bound on the estimator's variance consisting of only positive terms. This also reduces variance minimization in that context to the minimization of the $l^2$ norm of a certain matrix-vector product. We further show that if units are clustered, randomization can be done for each cluster individually.
 
Finally, we conclude and give an outlook discussing further research questions.

\chapter{Randomized Controlled Trials: Definition and Review} \label{sec:RCT-def-and-review}
This chapter aims to introduce RCTs, review their state-of-the-art in literature and discuss the idea of balance within subgroups both in terms of medical literature and the so-called Gram-Schmidt Walk Design. 

In more detail, we introduce RCTs formally (\Cref{sec:rct-formal-def}), give a summary of means of randomization from (medical) literature (\Cref{sec:types-of-randomization}), discuss a common way of approximating the so-called average treatment effect (\Cref{sec:approx-avg-trmt-effect}), give a statistical discussion of bias and variance (\Cref{sec:bias-vs-variance}) and finally review the Gram-Schmidt Walk Design algorithm and key theorems from Harshaw et al.\ (\Cref{sec:spectral-discussion,sec:gswd-main}).

\section{Formal Definition}\label{sec:rct-formal-def}

For consistency with the review of the Gram-Schmidt Walk Design, we stick to the notation of \cite{harshaw-BalancingCovariatesRandomized-2021}.

A randomized controlled trial is an experiment with $n$ participants, the so-called \textdef{units}. These units get partitioned into two groups: the \textdef{treatment} and \textdef{control} group, where they either receive a specific treatment or placebo. Because the placebo treatment can be seen as a ``null-treatment'', we refer for brevity to both groups as the \textdef{treatment groups}.\footnote{This is also closer to practice, where not only treatment versus placebo but also treatment A versus treatment B are compared} 

To which of the treatment groups a unit $i \in [n]$ gets assigned to is determined by the \textdef{assignment vector} $\vec z \in \{\pm 1\}^n$, where $z_i = +1$ means they are assigned to the treatment group $Z^+ := \{i \in [n]: z_i = +1 \}$. Analogously, $z_i = -1$ means they are assigned to the treatment group $Z^- := \{i \in [n]: z_i = -1 \}$.

Because the assignment $\vec z$ is random in a \emph{randomized} controlled trial, $\vec z$ is a random vector. The distribution of this random vector is called the \textdef{design}.

After the units have been treated, the experimenter measures a certain quantity for each unit, the \textdef{outcome}. We assume this to be a scalar value.
This outcome depends on a unit's treatment group: We measure $a_i$ if unit $i$ is in group $Z^+$. If it is in $Z^-$, we measure $b_i$. We refer to these quantities, that we potentially measure, as the \textdef{potential outcomes} $\vec{a}, \vec{b} \in \R^n$. Harshaw et al.\ introduce a vector combining both $\vec{a}, \vec b$: \textdef{\emph{the} potential outcomes vector} $\vec \mu :=\frac{1}{2} (\vec a + \vec b)$.

To assess the treatment's effect, we would ideally measure for each unit both $a_i$ and $b_i$. However, as the units are \emph{partitioned} into the treatment groups, we can either observe one of these quantities. This is encoded in the \textdef{observed outcome vector} $\vec y \in \R^n$:
\begin{equation}
	y_i := \begin{cases}
		a_i & \text{if } z_i = +1 \\
		b_i & \text{if } z_i = -1.
	\end{cases}
\end{equation}

The overall goal is to find the \textdef{average treatment effect} $\tau$:

\begin{equation}
	\tau = \frac{1}{n} \sum_{i \in [n]} a_i - b_i = \frac{1}{n} \langle \vec{1}, \vec{a} - \vec{b} \rangle.
\end{equation}

Because $a_i - b_i$ is unobservable,\footnote{We cannot measure both $a_i$ and $b_i$ for the same unit $i$, as it is either in group $Z^+$ or in group $Z^-$.} we use estimators $\hat\tau$ to approximate $\tau$. These estimators should, in expectation, match $\tau$. If this is the case, we say that $\hat\tau$ is \textdef{unbiased}. 

\paragraph*{Is unbiasedness enough?}
At first glance, a solely unbiased estimator could suffer from one issue: Even though it is unbiased, it could have a very high error $\abs{\hat\tau - \tau}$ in \textit{every} outcome. For example, consider an estimator that has the following error distribution:
\begin{equation}
	\hat\tau - \tau = \begin{cases}
		+1000 & \text{with probability }0.5 \\
		-1000 & \text{with probability }0.5
	\end{cases}
\end{equation}
In expectation over all possible designs, $\hat\tau$ is unbiased. But clearly, this is not a desirable estimator: The error is $\abs{\hat\tau - \tau} = 1000$ with certainty. We therefore want in addition the estimator to have small mean squared error
\begin{equation}
	\E{(\hat\tau -\tau)^2}.
\end{equation}

When taking a deeper look, we find that the mean squared error decreases for unbiased estimators when increasing the sample size $n$ (\Cref{sec:bias-vs-variance}). Therefore, solely unbiased estimators are widely considered a good tool. But in case of small study sizes, finding a design that minimizes an estimator's mean squared error is important.

We collect the definitions from above in the following table. This should serve as a reference point throughout this work.
\begin{center}
\begin{tabular}{  m{5.8cm} | m{5.8cm} } 

  Term & Definition \\ 
 \hline
 \hline
 	\textdef{Units} & the $n$ participants  \\ 
 \hline
 	\textdef{Potential outcomes} & $\vec a, \vec b$  \\ 
 \hline
 	\textdef{\emph{The} potential outcomes vector} & $\vec \mu :=\frac{1}{2} (\vec a + \vec b)$ \\
 \hline
 	\textdef{Observed outcomes} & $\vec y$  \\ 
 \hline
 	\textdef{Assignment} & $\vec z$  \\ 
  \hline
 	\textdef{Design} & distribution of assignments, i.e., distribution of random vector $\vec z$  \\ 
 \hline
 	\textdef{Treatment groups} &\specialcell{ $Z^+ := \{i \in [n]: z_i = +1 \}$,\\ $Z^- := \{i \in [n]: z_i = -1 \}$}  \\
 \hline
 	\textdef{Average treatment effect} &\rule{0pt}{3.5ex}$\tau = \frac{1}{n} \sum_{i \in [n]} a_i - b_i = \frac{1}{n} \langle \vec{1}, \vec{a} - \vec{b} \rangle$  \\ 
 
\end{tabular}
\end{center}

\section{Types of Randomization}{\label{sec:types-of-randomization}}

There are mainly four types of randomization used in practice: Complete randomization, random allocation, permuted-block randomization, and adaptive randomization \cite{chow-DesignAnalysisClinical-2013, kang-IssuesOutcomesResearch-2008}. We review them in the following.

\paragraph*{Complete randomization.}

The simplest form is \emph{complete randomization}, where $\Pr{z_i = 1} = \frac{1}{2}$, independent of any other unit $j$. We refer to this as the \textdef{i.i.d.\ randomization}. This is both simple in application and analysis, which is presumably the reason for its popularity. However, this randomization scheme might lead to unequal group sizes.

\paragraph*{Random allocation.}
To overcome the issue of different group sizes, \emph{random allocation} is used: $\frac{n}{2}$ units are drawn from the set of all $n$ units without replacement and assigned to (w.l.o.g.)\ the treatment group.

\paragraph*{Random allocation versus complete randomization.}

Intuition tells us that for large $n$, the difference between complete randomization and random allocation should be small. Therefore, it might be possible to use the easier-to-analyze complete randomization for large $n$. But how large should $n$ be, for being able to neglect the difference in group sizes?

A brief calculation allows for a quantitative argument, based on a simple estimation by Lachin \cite{lachin-PropertiesSimpleRandomization-1988}:

Let

\begin{equation*}
	S := \sum_{i \in [n]} z_i
\end{equation*}
be the sum of the assignment vector entries. $S=0$ indicates that the assignment is perfectly balanced. The sizes of $Z^+$,  $Z^-$ relate to $S$ in the following sense:

\begin{align*}
	|Z^+| &= \frac{n}{2} +\frac{S}{2}  \\
	|Z^-| &= \frac{n}{2} -\frac{S}{2} 
\end{align*}

A measure for imbalance, according to Lachin, is
\begin{equation*}
	\frac{\max\left(|Z^+|,|Z^-| \right)}{n} \stackrel{\text{ def. $S$ }}= \left|\frac{S}{2n} \right|+\frac{1}{2}.
\end{equation*}

Under complete randomization, linearity of expectation and the i.i.d\ property yields

\begin{align*}
	\mu &:= \E{S} = n \E{z_i} = 0 \\
	\sigma^2 &:= \Var{S} \stackrel{\text{indep.}}= \sum_{i \in [n]}\left (\underbrace{\E{z_i^2}}_{=1} - \underbrace{\E{z_i}^2}_{=0} \right) =  n.
\end{align*}
The central limit theorem applies for large enough $n$: \footnote{The CLT can be safely used for $n\geq 30$ \cite{lamorte-CentralLimitTheorem}, and our statement here will apply for larger $n$.}

\begin{align*}
	X := \frac{S-\mu}{\sigma} = \frac{S}{\sqrt{n}} \; \sim \; \mathcal{N}(0,1).
\end{align*}

Therefore, the probability that the imbalance is greater than $t\in[0,1]$ is:

\begin{align*}
	\Pr{\left|\frac{S}{2n} \right|+\frac{1}{2} > t} &\leq \Pr{ \frac{S}{\sqrt{n}} >2 \left(t-\frac{1}{2} \right)\sqrt{n} } +  \Pr{ \frac{S}{\sqrt{n}} < - 2 \left(t-\frac{1}{2} \right)\sqrt{n} }  \tag{union bound} \\
	&\approx 2 \Phi\left(2 \left(\frac{1}{2}-t \right)\sqrt{n} \right).  \tag{CLT}
\end{align*}

\begin{figure}[hbt]
  \centering
  \captionsetup{width=0.8\linewidth}
\begin{tikzpicture}

\definecolor{color0}{rgb}{0.83921568627451,0.152941176470588,0.156862745098039}

\begin{axis}[
height=9cm,
legend cell align={left},
legend style={fill opacity=0.8, draw opacity=1, text opacity=1, draw=white!80!black},
tick align=outside,
tick pos=left,
title={Number of units vs.\ imbalance},
width=12cm,
x grid style={white!69.0196078431373!black},
xlabel={Number of units \(\displaystyle n\)},
xmajorgrids,
xmin=-20, xmax=420,
xtick style={color=black},
y grid style={white!69.0196078431373!black},
ylabel={Probability of having imbalance \(\displaystyle >t\)},
ymajorgrids,
ymin=-0.0499334903921504, ymax=1.04999683287582,
ytick style={color=black}
]
\addplot [semithick, color0]
table {%
0 1
4.04040404040404 0.687672894255411
8.08080808080808 0.569671581587966
12.1212121212121 0.486234321388297
16.1616161616162 0.42137950374287
20.2020202020202 0.368688269361782
24.2424242424242 0.324755765402618
28.2828282828283 0.287495418869168
32.3232323232323 0.255508820294028
36.3636363636364 0.227799993988229
40.4040404040404 0.203627827093606
44.4444444444444 0.182422439451736
48.4848484848485 0.163734354324592
52.5252525252525 0.147201865212384
56.5656565656566 0.132529139996841
60.6060606060606 0.119470986771701
64.6464646464647 0.107821928159968
68.6868686868687 0.0974081616083543
72.7272727272727 0.0880815116621903
76.7676767676768 0.0797147932966756
80.8080808080808 0.0721981977016576
84.8484848484849 0.0654364338630845
88.8888888888889 0.0593464387919199
92.9292929292929 0.0538555223869285
96.969696969697 0.0488998492219099
101.010101010101 0.044423184850006
105.050505050505 0.04037585217168
109.090909090909 0.0367138563627041
113.131313131313 0.033398146339218
117.171717171717 0.0303939877735345
121.212121212121 0.027670427963097
125.252525252525 0.0251998368764741
129.292929292929 0.0229575117911992
133.333333333333 0.020921335337794
137.373737373737 0.0190714786444619
141.414141414141 0.0173901427629166
145.454545454545 0.015861332739773
149.49494949495 0.0144706596484004
153.535353535354 0.0132051666645675
157.575757575758 0.0120531758945961
161.616161616162 0.0110041531768688
165.656565656566 0.0100485884993831
169.69696969697 0.00917789002542132
173.737373737374 0.00838429001026202
177.777777777778 0.00766076113517946
181.818181818182 0.00700094198944863
185.858585858586 0.00639907060365385
189.89898989899 0.00584992508384362
193.939393939394 0.00534877052048203
197.979797979798 0.00489131145235933
202.020202020202 0.00447364925661177
206.060606060606 0.00409224391419709
210.10101010101 0.00374387966757986
214.141414141414 0.00342563414565487
218.181818181818 0.00313485058145306
222.222222222222 0.00286911279207661
226.262626262626 0.00262622262855756
230.30303030303 0.00240417963673231
234.343434343434 0.00220116269945111
238.383838383838 0.00201551345607066
242.424242424242 0.00184572131769626
246.464646464646 0.00169040991646165
250.505050505051 0.00154832484461341
254.545454545455 0.00141832255460523
258.585858585859 0.00129936030506489
262.626262626263 0.00119048704959754
266.666666666667 0.0010908351761253
270.707070707071 0.000999613014001702
274.747474747475 0.000916098034625738
278.787878787879 0.00083963067883791
282.828282828283 0.000769608751121125
286.868686868687 0.000705482326645895
290.909090909091 0.000646749122576476
294.949494949495 0.000592950289864902
298.989898989899 0.000543666586067046
303.030303030303 0.000498514893575204
307.070707070707 0.000457145051124714
311.111111111111 0.000419236969540987
315.151515151515 0.000384498005486852
319.191919191919 0.000352660569481904
323.232323232323 0.000323479946725845
327.272727272727 0.00029673231129314
331.313131313131 0.000272212916100306
335.353535353535 0.000249734442700796
339.393939393939 0.000229125496454361
343.434343434343 0.000210229233964629
347.474747474748 0.000192902110895108
351.515151515152 0.000177012739373323
355.555555555556 0.000162440845186801
359.59595959596 0.000149076315874022
363.636363636364 0.000136818331627392
367.676767676768 0.000125574571662341
371.717171717172 0.000115260489374373
375.757575757576 0.000105798650211039
379.79797979798 9.7118126734436e-05
383.838383838384 8.91539458475138e-05
387.878787878788 8.18465836089914e-05
391.919191919192 7.51415034715816e-05
395.959595959596 6.89887341503653e-05
400 6.33424836662397e-05
};
\addlegendentry{$t=0.6$}
\end{axis}

\end{tikzpicture}
  \caption{Probability of having an imbalance greater than $t=0.6$ as a function of the number of units, $n$. For $n  > 200$, this probability is negligibly small.}
  \label{fig:imbalancedness}
\end{figure}
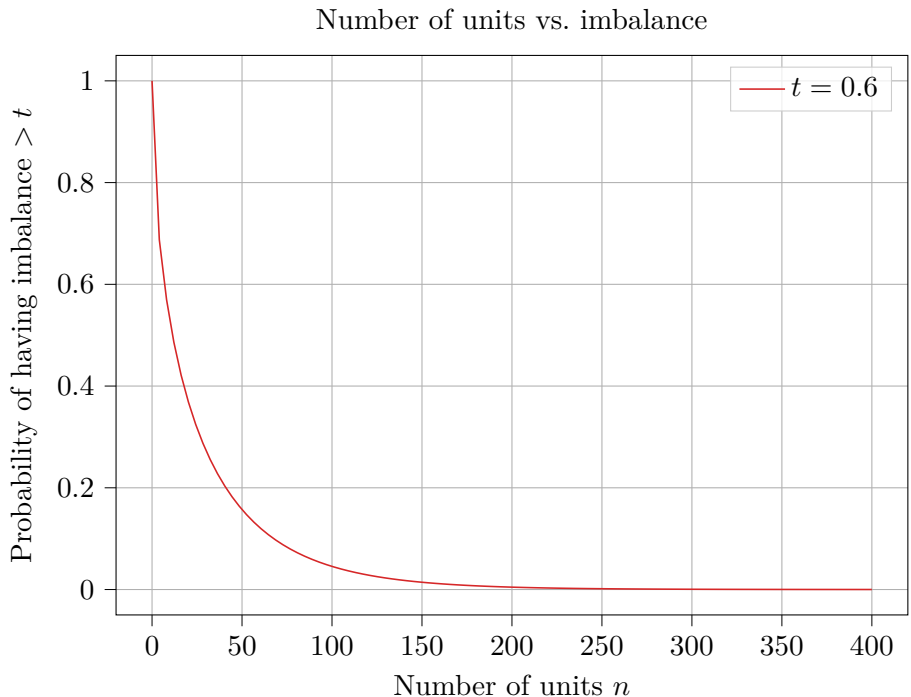

\Cref{fig:imbalancedness} shows this function for $t=0.6$. It can be seen that for $n>200$, the probability of having imbalance $>t=0.6$ \footnote{Imbalance $> t=0.6$ means that one group contains more than 60\% of all units.} is negligibly small ($< 0.0047$). This is the mathematical reason why practitioners use $n>200$ as a rule of thumb when imbalance is negligibly small and there is no reason to use random allocation. Instead, the easier-to-analyze method of complete randomization can be used \cite{kang-IssuesOutcomesResearch-2008,chow-DesignAnalysisClinical-2013}.

\begin{remark}[on notation]
	Harshaw et al.\ mention that random allocation is sometimes called complete randomization. However, this is not the case in standard literature such as \cite{chow-DesignAnalysisClinical-2013} or reviews such as \cite{lachin-PropertiesSimpleRandomization-1988,kang-IssuesOutcomesResearch-2008} where complete randomization clearly refers to the independent random allocation, as presented above. Therefore, it should be safe to use the term ``complete randomization'' for the i.i.d. design.
\end{remark} 

What if both treatment and control group are balanced, but subgroups are not? This is unfavorable, as subgroups could be predictive of the outcome. In medical trials, this is referred to as ``prognostic factors'' if, e.g., the age influences the observed outcome. To overcome this issue, the portion of treatment / control assignments gets balanced \emph{within} each subgroup (also known as stratum). This technique is referred to as \textdef{stratified randomization}. The following randomization schemes are both stratified randomization techniques.

\paragraph*{Permuted-Block Randomization.} \label{sec:permuted-block-randomization}
Units get partitioned into blocks, and randomization is performed within each block, commonly using the \emph{random allocation} scheme. For example, units might sequentially enter a study over an extended period of time (e.g., patients in a hospital). To ensure that any covariates that change over time are balanced, units get allocated into blocks according to their time of entry. Random allocation is now performed for each block, thereby ensuring balancedness of time-fluctuating covariates.

\paragraph*{Adaptive Randomization.}
In adaptive randomization, the probability of getting assigned to treatment / control gets adapted over time, based on covariates measured until the time of entry. This is highly situation-dependent, and therefore does not allow for a good general analysis.

\vspace{1em}
In \Cref{sec:spectral-discussion,sec:gswd-main}, we will review a paper \cite{harshaw-BalancingCovariatesRandomized-2021} that performs at its very heart stratification -- a technique used for decades.\footnote{For a detailed review of stratified randomization and its usage since 1966, see \cite{kernan-StratifiedRandomizationClinical-1999} } The paper's key contribution is that it provides an algorithm with good bounds on the variance of a common average treatment effect estimator, which is defined in the following.

\section{Approximating the Average Treatment Effect} \label{sec:approx-avg-trmt-effect}

A widely used estimator of the average treatment effect is the Horvitz-Thompson estimator \cite{horvitz-GeneralizationSamplingReplacement-1952}. In the most general case, the Horvitz-Thompson estimator is defined as
\begin{equation}
	\tauht = \frac{1}{n} \left( \sum_{i \in Z^+} \frac{y_i}{\Pr{z_i = +1}} - \sum_{i \in Z^-} \frac{y_i}{\Pr{z_i = -1}}  \right).
\end{equation}

For the remainder of this work, we will assume $\Pr{z_i = +1} = \Pr{z_i = -1} = \frac{1}{2}$ for all units, as this gives -- in expectation -- equal group sizes.\footnote{This assumption also follows Harshaw et al. All calculations can be generalized by taking the broader definition of the Horvitz-Thompson estimator, which includes the group assignment probabilities.} In this case, we get a simpler expression:

\begin{definition}[Horvitz-Thompson estimator]\label{def:tauht} For a design with $\Pr{z_i = +1} = \Pr{z_i = -1} = \frac{1}{2}$ for all $i \in [n]$, the Horvitz-Thompson estimator is
\begin{equation*}
	\tauht = \frac{2}{n}\langle \vec{z}, \vec{y} \rangle.
\end{equation*}
\end{definition}

 \begin{lemma}\label{lem:tauht-unbiased}
  $\tauht$ is unbiased for a design with $\Pr{z_i = +1} = \Pr{z_i = -1} = \frac{1}{2}$ for all $i \in [n]$.
 \end{lemma}

\begin{proof}
	\begin{align*}
		\E{\tauht} 
		&= \E{\frac{2}{n}\langle \vec{z}, \vec{y} \rangle} \tag{by Definition \ref{def:tauht}} \\
		&= \frac{2}{n}\sum_{i \in [n]} \E{z_i y_i} \tag{by linearity of $\mathbb{E}$}\\
		&= \frac{2}{n}\sum_{i \in [n]} \frac{1}{2} a_i -  \frac{1}{2} b_i  \tag{by definition of $\vec y$}\\
		&= \frac{1}{n} \langle \vec 1 , \vec a - \vec b \rangle \\
		&= \tau.
	\end{align*}
	
\end{proof}

As $\Var{}{\tauht} = \E{}{\left(\tauht - \E{z}{\tauht}\right)^2}$, the unbiasedness (\Cref{lem:tauht-unbiased}) implies:
\begin{corollary}
The variance of $\tauht$ is equal to the mean squared error
	\begin{equation}
		\Var{\tauht} = \E{\left( \tauht - \tau \right)^2}.
	\end{equation}
\end{corollary}
We will therefore use the term \emph{variance of the estimator} in the following.
It shall be noted that in medical literature, the term ``increasing precision'' is sometimes used. Mathematically, this means reducing the variance.

Note that unbiasedness or low variance always refers to a \emph{combination} of both design and estimator.

\section{Bias Versus Variance -- Which one Decreases Naturally?} \label{sec:bias-vs-variance}

 Talks with experimenters showed that they clearly prefer an unbiased estimator-design combination over a low variance one, noting that a low variance can also be achieved by simply increasing the experiment size $n$. Here, we give a mathematical justification for this attitude, at the example of a slightly modified estimator based on $\tauht$. 
 
 Let 
 \begin{equation}
 	\hat\tau := \frac{1}{n} \langle \vec z, \vec y \rangle 
 \end{equation}
 be our estimator.
 
 Suppose, for simplicity, that each unit has the same $\E{z_i y_i}= \frac{1}{2}\left(a_i-b_i\right) =:  \frac{1}{2} c$ and $\Var{z_i y_i} =  \frac{1}{2}\left( a_i -c \right)^2 +  \frac{1}{2}\left( b_i + c \right)^2=:s$. Therefore, the expectation of $\hat\tau$ is
 \begin{equation}
 	\E{\hat\tau} \stackrel{\text{def.}}{=} \E{\frac{1}{n}\sum_{i \in [n]} z_i y_i}  \stackrel{\substack{\text{linearity of $\mathbb{E}$},\\ \mathbb{E}[z_i y_i]}}= \frac{1}{n} \left( n \cdot \frac{1}{2} c \right) = \frac{1}{2} c.
 \end{equation}
 
 which is biased, as the average treatment effect is
 \begin{equation}
 	\tau = \frac{1}{n} \sum_{i \in [n]} a_i - b_i = c.
 \end{equation}
 
 But the variance under the i.i.d.\ design is 
 \begin{equation} \label{eq:variance-decreases-with-high-n}
 	\Var{\hat\tau} = \Var{\frac{1}{n} \sum_{i \in [n]} z_i y_i} \stackrel{\text{indep.}}{=}\frac{1}{n^2} n \cdot s = \frac{1}{n} s.
 \end{equation}
 The key observation here is the factor $\frac{1}{n}$, which exists in the final variance term, but not in the final bias term:
By choosing the i.i.d.\ design, the variance decreases with $\frac{1}{n}$, but the expectation is fixed independent of $n$. Therefore, a biased estimator cannot be ``made unbiased'' by increasing $n$, but the precision of a high-variance estimator can be increased by increasing $n$. 

Note that besides the i.i.d.\ design, we used in Equation \ref{eq:variance-decreases-with-high-n} the SUTVA (the $y_i$ do not depend on each other). Therefore, such a calculation becomes more challenging in the presence of spillover effects (peer influence). We will pick up this idea in \Cref{sec:network-effects}.

\section{Spectral Discussion} \label{sec:spectral-discussion}

In this section, we first find a simple expression for the error of $\tauht$, subsequently analyze the variance by using spectral decomposition, and finally draw conclusions about a tradeoff between potential performance and robustness of $\tauht$. This insight will be key for realizing the possibilities and limitations of a good design.

\begin{lemma}[Error of $\tauht$] \label{lem:error-of-tau-ht}
	The error of $\tauht$ is:
	\begin{equation*}
		\tauht - \tau = \frac{2}{n} \langle \vec z, \vec \mu \rangle
	\end{equation*}
\end{lemma} 
	
\begin{proof}
	\begin{align}
		\tauht - \tau & \stackrel{\text{def.}}{=} \frac{2}{n}\expval{\vec z, \vec y}- \expval{\vec 1, \vec a-\vec b} \\
		&= \frac{1}{n} \left( \sum_{i \in Z^+} \left( 2a_i-(a_i-b_i) \right) -  \sum_{i \in Z^-} \left( 2b_i+(a_i-b_i) \right) \right) \\
		&= \frac{1}{n} \left( \sum_{i \in Z^+} (a_i+b_i) -  \sum_{i \in Z^-} (a_i + b_i)   \right) \\
		&= \frac{2}{n} \expval{\vec z, \vec \mu}
\end{align}
\end{proof}

This lets us rewrite the variance:

\begin{equation}
	\Var{\tauht} = \E{\left( \tauht - \tau \right)^2} \stackrel{\text{\Cref{lem:error-of-tau-ht}}}{=} \E{\frac{4}{n^2} \vec \mu' \vec z \vec z' \vec \mu} = \frac{4}{n^2} \vec \mu' \E{\vec z \vec z'} \vec \mu \stackrel{(*)}{=} \frac{4}{n^2} \vec \mu' \Cov{\vec z} \vec \mu 
\end{equation} 

where $(*)$ follows from the fact that $\E{\vec z} = \vec 0$.

As $\Cov{\vec z}$ is a symmetric matrix, the spectral theorem applies, and we can eigendecompose:
\begin{equation}
	\Cov{\vec z} = \mat V \mat \Lambda \mat V' 
\end{equation}
	
where $\mat \Lambda=\operatorname{diag}(\lambda_i)$  consists of the eigenvalues and $\mat V$ of the eigenvectors of $\Cov{\vec v}$. According to the spectral theorem, $\mat V$ forms a basis of $\R^n$. We can therefore express $\vec \mu$ in terms of the eigenbasis $\mat V$:

\begin{equation}
	\exists \vec w \in \R^n: \; \frac{\vec \mu}{\norm{\vec \mu}_2} = \mat V \vec w \quad \text{with }\norm{\vec w}_2 = 1.
\end{equation}

We thus get for the variance:
\begin{equation} \label{eq:variance-spectral-expression}
	\Var{\tauht} = \frac{4}{n^2} \vec \mu' \Cov{\vec z} \vec \mu =  \frac{4}{n^2} \norm{\vec \mu}_2^2 \vec w' \mat V' \mat V \mat \Lambda \mat V' \mat V \vec w = \frac{4 \norm{\vec \mu}_2^2}{n^2} \underbrace{\sum_{i \in [n]} w_i^2 \lambda_i}_{(*)}
\end{equation} 
where $(*)$ is a convex combination of the eigenvalues of $\Cov{\vec z}$, since $\norm{\vec w}_2 = 1$.

This gives a first insight for designing good experiments: If possible, we should align the smallest eigenvector of $\Cov{\vec z}$ with $\vec \mu$ while making the biggest eigenvectors orthogonal to $\vec \mu$. However, we do not know $\vec \mu$ (as it directly depends on both potential outcomes $\vec a$, $\vec b$). Nonetheless, we might be able to predict $\vec \mu$ based on some covariate information: This is the topic of \Cref{sec:gswd-main}.

Because Equation \ref{eq:variance-spectral-expression} contains a convex combination of the eigenvectors, we can turn it into a bound on the worst-case variance of a design:
\begin{lemma}\label{lem:worst-case-variance-of-a-design}
	The worst-case variance of $\tauht$ can be bounded by $\lambda_{max}$, the maximal eigenvalue of $\Cov{\vec z}$:
	\begin{equation*}
		\max_{\vec \mu \in \R^n} \left( \frac{1}{\norm{\vec \mu}_2^2}  \Var{\tauht} \right) = \frac{4}{n^2} \lambda_{max}
	\end{equation*}
	where worst-case refers to the alignment of the potential outcomes vector $\vec \mu$ with $\Cov{\vec z}$ that gives maximal (i.e.\ worst) variance.
\end{lemma} 

\begin{lemma} \label{lem:sum-eigen-vals-covz}
	For the sum of the eigenvalues of $\Cov{\vec z}$, we have:
	\begin{equation*}
		\sum_{i \in [n]} \lambda_i = n
	\end{equation*}
\end{lemma}

\begin{proof}
	For any $i \in [n]$, we have:
	\begin{equation*}
		\Cov{z}_{i,i} = \Var{z_i} =\Pr{z_i = +1} \cdot (+1)^2 + \Pr{z_i = -1} \cdot (-1)^2 = 1
	\end{equation*}
	And as the sum of a matrix's eigenvalues is its trace, we have 
	\begin{equation}
		\sum_{i \in [n]} \lambda_i = \tr{\Cov{z}} = n
	\end{equation}
\end{proof}

This shows an inherent trade-off between a design's robustness and potential performance: We see in Equation \ref{eq:variance-spectral-expression} that $\lambda_{min}$ determines how good a design's potential performance is, while $\lambda_{max}$ expresses how bad a design's worst-case variance can get.\footnote{If the potential outcome vector $\vec \mu$ is aligned with the smallest eigenvector, we get a low variance. If it is aligned with the biggest eigenvector, we get a high variance.} The best design would thus have both a small $\lambda_{max}$ (robust against bad $\vec \mu$) and a small $\lambda_{min}$ (good best-case performance). However, \Cref{lem:sum-eigen-vals-covz} shows that $\sum_{i \in [n]} \lambda_i = n$, so that we cannot have both small $\lambda_{min}$ and $\lambda_{max}$.

The i.i.d. design with $\lambda_1 = \ldots = \lambda_n = 1$ has minimal $\lambda_{max}$ so that it is robust. However, it has no good potential performance.
A design with $\lambda_{min}<1$, on the other hand, has good potential performance (for $\vec \mu$ being aligned with the minimal eigenvector) but bad worst-case performance, as $\lambda_{max}>1$.

Harshaw et al.\ use this argument to state that no design can be uniformly better than all others. There will always exist potential outcomes $\vec \mu$ where one design has better (i.e., lower) variance than the other.
If we know that $\vec \mu$ lies predominantly in a specific direction, we can create good designs. This is the baseline assumption of the Gram-Schmidt Walk Design and is covered in the next section.

\section{The Gram-Schmidt Walk Design \cite{harshaw-BalancingCovariatesRandomized-2021}} \label{sec:gswd-main}
As we saw in the previous section, we cannot improve the i.i.d.\ design without making further assumptions on the potential outcomes. 

The assumption made in \cite{harshaw-BalancingCovariatesRandomized-2021} is that there exists some information (\emph{covariate information}) for each unit that is predictive of the potential outcome vector $\vec\mu$. As shown in the introduction, this idea is not new and has been discussed under the term \emph{stratification} in the clinical trial literature at least since the 1970s \cite{zelen-RandomizationStratificationPatients-1974}.

However, a key contribution from Harshaw et al. is to give an algorithm under these assumptions with good bounds on the variance. After stating the assumptions formally, the goal of this section is to both give intuition behind the algorithm (\Cref{fig:gswd-intuition}) as well as review some of the main theorems.

\subsection{Assumptions on Potential Outcomes} \label{sec:gswd-assumptions-on-potential-outcomes}
Formally, we represent the $d$ covariates we have for each unit in the \emph{covariate matrix} $\mat X \in \mathbb{R}^{n,d}$, where the $i$th row represents the covariates of unit $i$. The covariates are linearly predictive of the potential outcomes if $\vec\mu$ is close to the column span of $\mat X$:
\begin{equation}
	\vec \mu \approx \mat X \vec \beta \quad \text{for some vector } \beta \in \mathbb{R}^d.
\end{equation}

If we denote by $\vec\beta$ the vector representing the linear function between covariates and potential outcomes best 

\begin{equation}
	\vec\beta := \argmin_{\tilde\beta \in \mathbb{R}^d} \norm{\vec\mu -\mat X \tilde\beta}_2,
\end{equation}

we can decompose $\vec \mu$ into $\vec\mu = \vec{\hat{\mu}} + \vec\epsilon$ where $\vec{\hat{\mu}} := \mat X \vec\beta$, $\vec\epsilon := \vec\mu - \vec{\hat{\mu}}$ and get for the variance of the Horvitz-Thompson estimator:

\begin{equation}
	\Var{\tauht} = \vec\mu' \Cov{\vec z}\vec \mu = \vec{\hat{\mu}}' \Cov{\vec z} \vec{\hat{\mu}} + 2 \vec{\hat{\mu}}' \Cov{\vec z} \vec\epsilon + \vec \epsilon'   \Cov{\vec z} \vec \epsilon
\end{equation}

To minimize that expression, we would like to align the smallest eigenvector of $\Cov{\vec z}$ with both $ \vec{\hat{\mu}}$ and $ \vec\epsilon$. Since they are orthogonal, however, we cannot align to both vectors at the same time.
Nonetheless, by our assumption that $\mat X$ is predictive of $\vec \mu$, we can assume that $\vec \epsilon$ is small and thus  

\begin{equation}
	\Var{\tauht} \approx \vec{\hat{\mu}}' \Cov{\vec z} \vec{\hat{\mu}}
\end{equation}

As $ \vec{\hat{\mu}} \in \operatorname{Span}(\mat X)$, we can further simplify:
\begin{equation}
	\vec{\hat{\mu}}' \Cov{\vec z} \vec{\hat{\mu}} = \vec \beta' \mat X' \Cov{\vec z}\mat X \vec \beta = \vec \beta ' \Cov{\mat X' \vec z} \vec \beta.
\end{equation}

If the covariates are predictive, we should therefore focus on $ \Cov{\mat X' \vec z}$ rather than solely on $\Cov{\vec z}$.

\vspace{\baselineskip}
In the following sections, we first give an intuition for the Gram-Schmidt Walk Design algorithm (\Cref{sec:gswd-intuition}), then describe the algorithm itself and establish some mathematical properties (\Cref{sec:gswd-algo}) and finally analyze the Horvitz-Thompson estimator's error under that design, corresponding to the bias (\Cref{sec:gswd-analysis-bias}) and mean square error (\Cref{sec:gswd-analysis-variance}).

\subsection{Intuition} \label{sec:gswd-intuition}
The overall goal is to randomly find an assignment vector $\vec z \in \{ \pm 1\}^n$, that assigns each of the $n$ elements into either $Z^+$ or $Z^-$.

There is an inherent tradeoff between robustness, achieved by i.i.d.\ randomization, and potential performance,\footnote{The more predictive the covariates are for the potential outcomes, the better the \emph{potential} performance} achieved by balancing the covariates between both $Z^+$ and $Z^-$ groups (\Cref{sec:spectral-discussion}). This founds the necessity for a trade-off parameter $\Phi$ between robustness ($\Phi = 1$) and potential performance ($\Phi = 0$), which needs to be specified by the experimenter. 

How can we connect (1) i.i.d. randomization and (2) balance of covariates in order to find a randomized design corresponding to a given $\Phi \in (0,1]$?

Intuitively, (2) corresponds to balancing some measure of the $n$ covariate vectors. 
We can phrase (1) in terms of (2): i.i.d. random assignment of $n$ elements into two groups $Z^+, Z^-$ is the same as randomly balancing $n$ orthogonal vectors: It is impossible to balance these so that the best random balance is just an i.i.d. randomization. Let us use the unit vectors $e_i$ as orthogonal vectors.

 Combining robustness and balance, we, therefore, aim to balance $n$ vectors consisting of both an orthogonal part $e_i$, scaled by $\approx \Phi$, and a part consisting of the $i$th covariate vector $\mat X_{i,:}$, scaled by $\approx 1-\Phi$. 
 
 Harshaw et al.\ achieve this by balancing the column vectors of

\begin{equation}
	\mat B = \begin{bmatrix}
		\sqrt{\Phi} \mat I \\
		\xi^{-1} \sqrt{1-\Phi} \mat X'
	\end{bmatrix}
	\quad \in \R^{d+n,n}
\end{equation}

where $\xi = \max_{i \in [n]} \norm{\mat X_{i,:} }_2$ is some scaling factor.\footnote{ $\xi$ is only important in a very detailed proof. We can think of it as some given scaling factor.} Consider the extreme cases: For $\Phi = 1$, $\mat B$ consists of only orthogonal unit vectors, which cannot be balanced. So the best way to balance will be an i.i.d. assignment. If $\Phi = 0$, $\mat B$ consists of only the covariate vectors. Thus, balancing the column vectors of $\mat B$ corresponds to balancing the covariate vectors.

In summary, the goal is to randomly find a vector $\vec z$ on one of the corners of the hypercube $\{ \pm 1\}^n$, while somehow balancing the columns of $\mat B$.

\begin{figure}[h!]
     \centering
     \begin{subfigure}[b]{0.3\textwidth}
         \centering
         \footnotesize
         \def\svgwidth{\textwidth}
\begingroup%
  \makeatletter%
  \providecommand\color[2][]{%
    \errmessage{(Inkscape) Color is used for the text in Inkscape, but the package 'color.sty' is not loaded}%
    \renewcommand\color[2][]{}%
  }%
  \providecommand\transparent[1]{%
    \errmessage{(Inkscape) Transparency is used (non-zero) for the text in Inkscape, but the package 'transparent.sty' is not loaded}%
    \renewcommand\transparent[1]{}%
  }%
  \providecommand\rotatebox[2]{#2}%
  \newcommand*\fsize{\dimexpr\f@size pt\relax}%
  \newcommand*\lineheight[1]{\fontsize{\fsize}{#1\fsize}\selectfont}%
  \ifx\svgwidth\undefined%
    \setlength{\unitlength}{425.19685039bp}%
    \ifx\svgscale\undefined%
      \relax%
    \else%
      \setlength{\unitlength}{\unitlength * \real{\svgscale}}%
    \fi%
  \else%
    \setlength{\unitlength}{\svgwidth}%
  \fi%
  \global\let\svgwidth\undefined%
  \global\let\svgscale\undefined%
  \makeatother%
  \begin{picture}(1,1)%
    \lineheight{1}%
    \setlength\tabcolsep{0pt}%
    \put(0,0){\includegraphics[width=\unitlength,page=1]{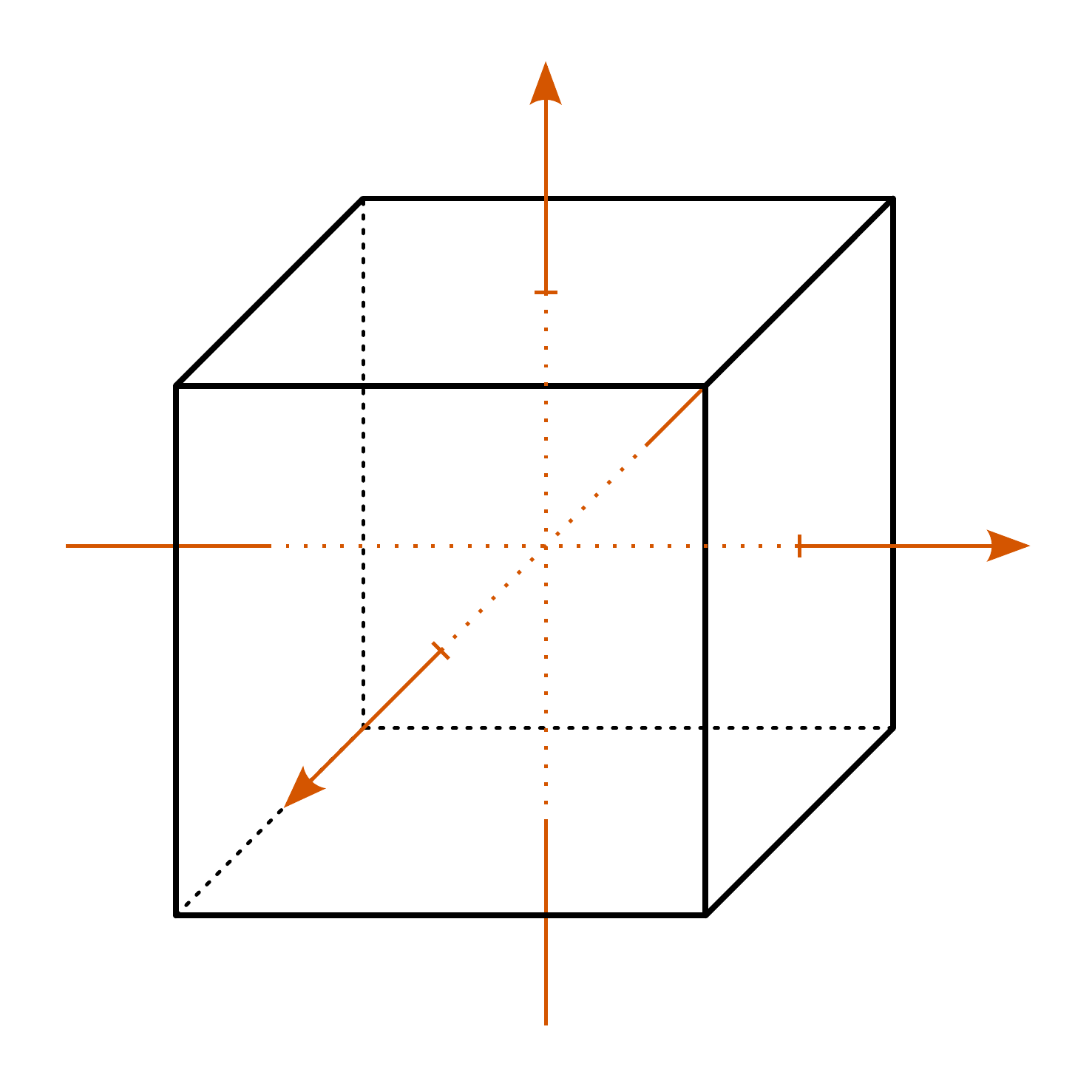}}%
    \put(0.52837723,0.72586047){\makebox(0,0)[lt]{\smash{\begin{tabular}[t]{l}$1$\end{tabular}}}}%
    \put(0.72723994,0.44442952){\makebox(0,0)[lt]{\smash{\begin{tabular}[t]{l}$1$\end{tabular}}}}%
    \put(0.37132678,0.4290388){\makebox(0,0)[lt]{\smash{\begin{tabular}[t]{l}$1$\end{tabular}}}}%
    \put(0,0){\includegraphics[width=\unitlength,page=2]{gswd_visual_1.pdf}}%
    \put(0.51863113,0.46058843){\makebox(0,0)[lt]{\smash{\begin{tabular}[t]{l}$\vec z_0$\end{tabular}}}}%
  \end{picture}%
\endgroup%

         \caption{}
     \end{subfigure}
     \hfill
     \begin{subfigure}[b]{0.3\textwidth}
         \centering
         \footnotesize
         \def\svgwidth{\textwidth}
\begingroup%
  \makeatletter%
  \providecommand\color[2][]{%
    \errmessage{(Inkscape) Color is used for the text in Inkscape, but the package 'color.sty' is not loaded}%
    \renewcommand\color[2][]{}%
  }%
  \providecommand\transparent[1]{%
    \errmessage{(Inkscape) Transparency is used (non-zero) for the text in Inkscape, but the package 'transparent.sty' is not loaded}%
    \renewcommand\transparent[1]{}%
  }%
  \providecommand\rotatebox[2]{#2}%
  \newcommand*\fsize{\dimexpr\f@size pt\relax}%
  \newcommand*\lineheight[1]{\fontsize{\fsize}{#1\fsize}\selectfont}%
  \ifx\svgwidth\undefined%
    \setlength{\unitlength}{425.19685039bp}%
    \ifx\svgscale\undefined%
      \relax%
    \else%
      \setlength{\unitlength}{\unitlength * \real{\svgscale}}%
    \fi%
  \else%
    \setlength{\unitlength}{\svgwidth}%
  \fi%
  \global\let\svgwidth\undefined%
  \global\let\svgscale\undefined%
  \makeatother%
  \begin{picture}(1,1)%
    \lineheight{1}%
    \setlength\tabcolsep{0pt}%
    \put(0,0){\includegraphics[width=\unitlength,page=1]{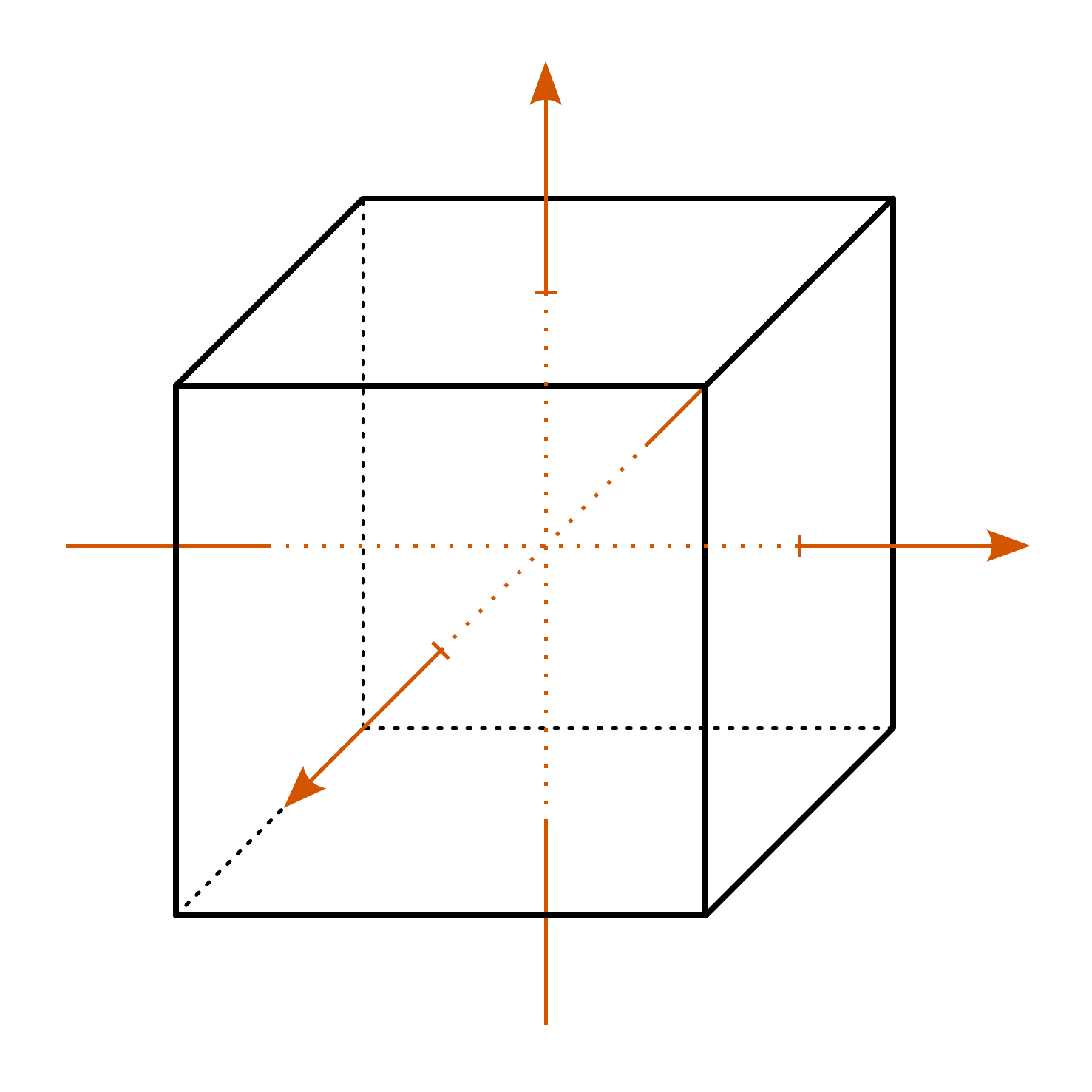}}%
    \put(0.52837723,0.72586047){\makebox(0,0)[lt]{\smash{\begin{tabular}[t]{l}$1$\end{tabular}}}}%
    \put(0.72723994,0.44442952){\makebox(0,0)[lt]{\smash{\begin{tabular}[t]{l}$1$\end{tabular}}}}%
    \put(0.37132678,0.4290388){\makebox(0,0)[lt]{\smash{\begin{tabular}[t]{l}$1$\end{tabular}}}}%
    \put(0,0){\includegraphics[width=\unitlength,page=2]{gswd_visual_2.pdf}}%
    \put(0.53660161,0.60666082){\makebox(0,0)[lt]{\smash{\begin{tabular}[t]{l}$\vec u$\end{tabular}}}}%
  \end{picture}%
\endgroup%

         \caption{}
     \end{subfigure}
     \hfill
     \begin{subfigure}[b]{0.3\textwidth}
         \centering
         \footnotesize
         \def\svgwidth{\textwidth}
\begingroup%
  \makeatletter%
  \providecommand\color[2][]{%
    \errmessage{(Inkscape) Color is used for the text in Inkscape, but the package 'color.sty' is not loaded}%
    \renewcommand\color[2][]{}%
  }%
  \providecommand\transparent[1]{%
    \errmessage{(Inkscape) Transparency is used (non-zero) for the text in Inkscape, but the package 'transparent.sty' is not loaded}%
    \renewcommand\transparent[1]{}%
  }%
  \providecommand\rotatebox[2]{#2}%
  \newcommand*\fsize{\dimexpr\f@size pt\relax}%
  \newcommand*\lineheight[1]{\fontsize{\fsize}{#1\fsize}\selectfont}%
  \ifx\svgwidth\undefined%
    \setlength{\unitlength}{425.19685039bp}%
    \ifx\svgscale\undefined%
      \relax%
    \else%
      \setlength{\unitlength}{\unitlength * \real{\svgscale}}%
    \fi%
  \else%
    \setlength{\unitlength}{\svgwidth}%
  \fi%
  \global\let\svgwidth\undefined%
  \global\let\svgscale\undefined%
  \makeatother%
  \begin{picture}(1,1)%
    \lineheight{1}%
    \setlength\tabcolsep{0pt}%
    \put(0,0){\includegraphics[width=\unitlength,page=1]{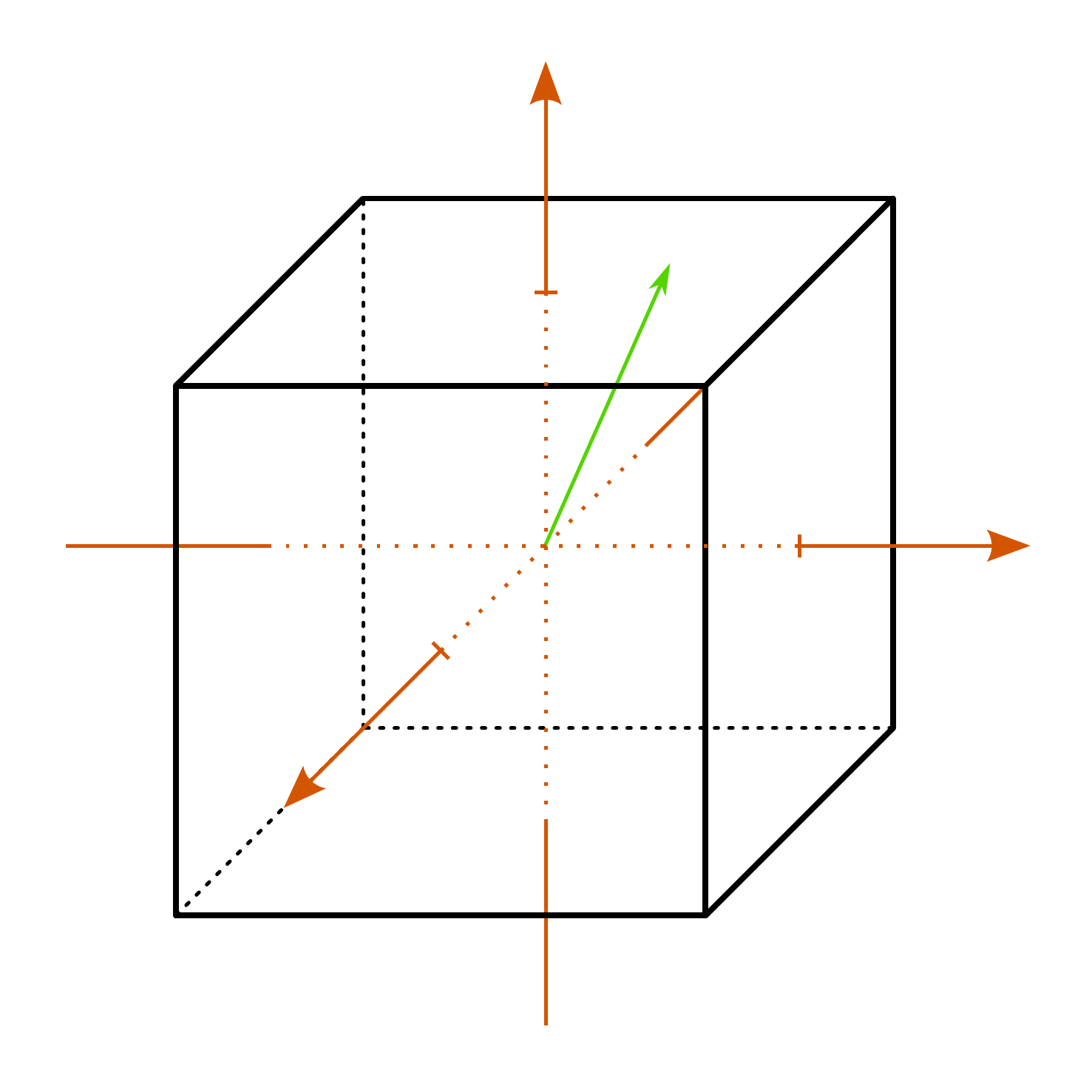}}%
    \put(0.52837723,0.72586047){\makebox(0,0)[lt]{\smash{\begin{tabular}[t]{l}$1$\end{tabular}}}}%
    \put(0.72723994,0.44442952){\makebox(0,0)[lt]{\smash{\begin{tabular}[t]{l}$1$\end{tabular}}}}%
    \put(0.37132678,0.4290388){\makebox(0,0)[lt]{\smash{\begin{tabular}[t]{l}$1$\end{tabular}}}}%
    \put(0,0){\includegraphics[width=\unitlength,page=2]{gswd_visual_3.pdf}}%
  \end{picture}%
\endgroup%

         \caption{}
     \end{subfigure} \\[1em]
      \begin{subfigure}[b]{0.3\textwidth}
         \centering
         \footnotesize
         \def\svgwidth{\textwidth}
\begingroup%
  \makeatletter%
  \providecommand\color[2][]{%
    \errmessage{(Inkscape) Color is used for the text in Inkscape, but the package 'color.sty' is not loaded}%
    \renewcommand\color[2][]{}%
  }%
  \providecommand\transparent[1]{%
    \errmessage{(Inkscape) Transparency is used (non-zero) for the text in Inkscape, but the package 'transparent.sty' is not loaded}%
    \renewcommand\transparent[1]{}%
  }%
  \providecommand\rotatebox[2]{#2}%
  \newcommand*\fsize{\dimexpr\f@size pt\relax}%
  \newcommand*\lineheight[1]{\fontsize{\fsize}{#1\fsize}\selectfont}%
  \ifx\svgwidth\undefined%
    \setlength{\unitlength}{425.19685039bp}%
    \ifx\svgscale\undefined%
      \relax%
    \else%
      \setlength{\unitlength}{\unitlength * \real{\svgscale}}%
    \fi%
  \else%
    \setlength{\unitlength}{\svgwidth}%
  \fi%
  \global\let\svgwidth\undefined%
  \global\let\svgscale\undefined%
  \makeatother%
  \begin{picture}(1,1)%
    \lineheight{1}%
    \setlength\tabcolsep{0pt}%
    \put(0,0){\includegraphics[width=\unitlength,page=1]{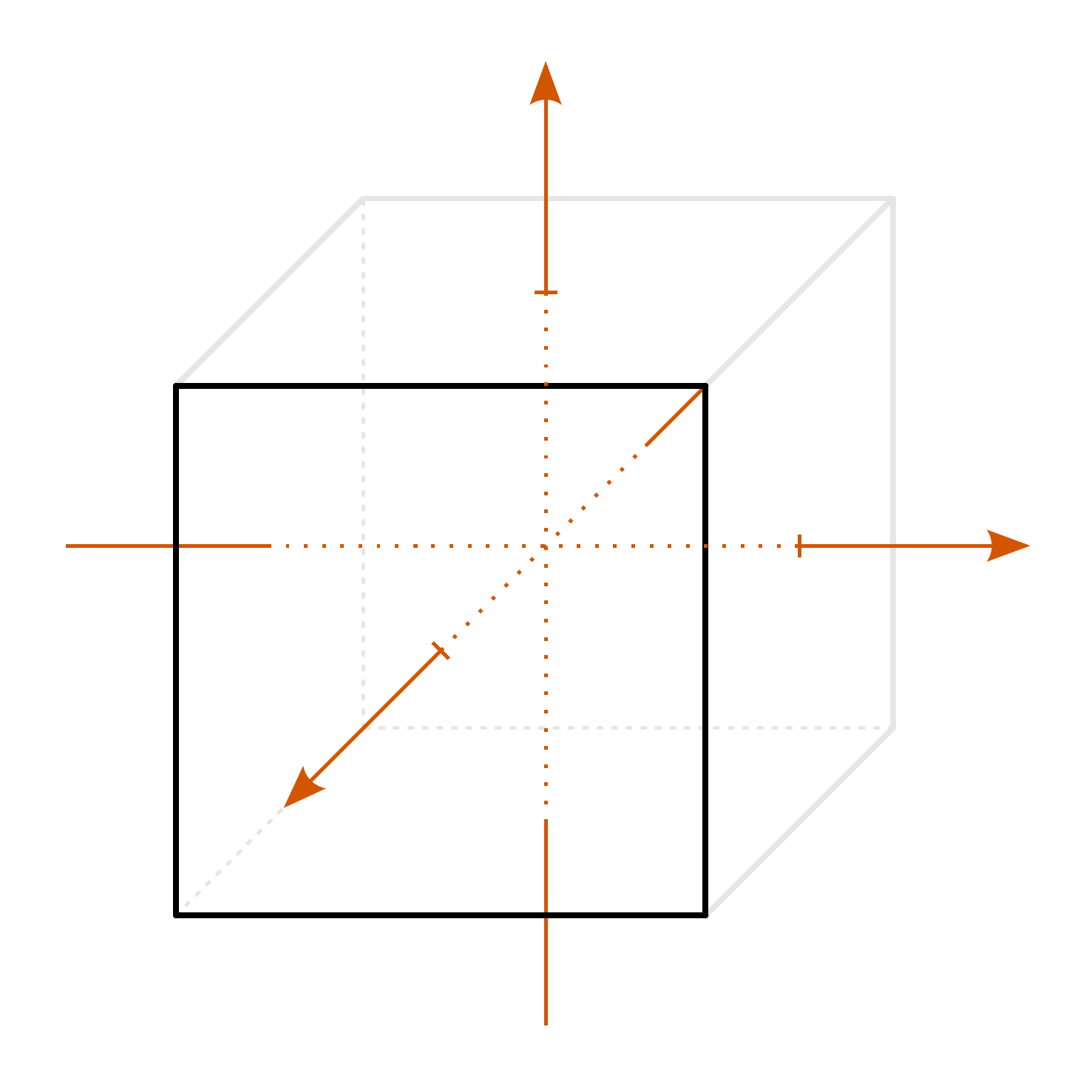}}%
    \put(0.52837723,0.72586047){\makebox(0,0)[lt]{\smash{\begin{tabular}[t]{l}$1$\end{tabular}}}}%
    \put(0.72723994,0.44442952){\makebox(0,0)[lt]{\smash{\begin{tabular}[t]{l}$1$\end{tabular}}}}%
    \put(0.37132678,0.4290388){\makebox(0,0)[lt]{\smash{\begin{tabular}[t]{l}$1$\end{tabular}}}}%
    \put(0,0){\includegraphics[width=\unitlength,page=2]{gswd_visual_4.pdf}}%
    \put(0.410927,0.21906611){\makebox(0,0)[lt]{\smash{\begin{tabular}[t]{l}$\vec z_1$\end{tabular}}}}%
  \end{picture}%
\endgroup%

         \caption{}
     \end{subfigure}
     \hfill
     \begin{subfigure}[b]{0.3\textwidth}
         \centering
         \footnotesize
         \def\svgwidth{\textwidth}
\begingroup%
  \makeatletter%
  \providecommand\color[2][]{%
    \errmessage{(Inkscape) Color is used for the text in Inkscape, but the package 'color.sty' is not loaded}%
    \renewcommand\color[2][]{}%
  }%
  \providecommand\transparent[1]{%
    \errmessage{(Inkscape) Transparency is used (non-zero) for the text in Inkscape, but the package 'transparent.sty' is not loaded}%
    \renewcommand\transparent[1]{}%
  }%
  \providecommand\rotatebox[2]{#2}%
  \newcommand*\fsize{\dimexpr\f@size pt\relax}%
  \newcommand*\lineheight[1]{\fontsize{\fsize}{#1\fsize}\selectfont}%
  \ifx\svgwidth\undefined%
    \setlength{\unitlength}{425.19685039bp}%
    \ifx\svgscale\undefined%
      \relax%
    \else%
      \setlength{\unitlength}{\unitlength * \real{\svgscale}}%
    \fi%
  \else%
    \setlength{\unitlength}{\svgwidth}%
  \fi%
  \global\let\svgwidth\undefined%
  \global\let\svgscale\undefined%
  \makeatother%
  \begin{picture}(1,1)%
    \lineheight{1}%
    \setlength\tabcolsep{0pt}%
    \put(0,0){\includegraphics[width=\unitlength,page=1]{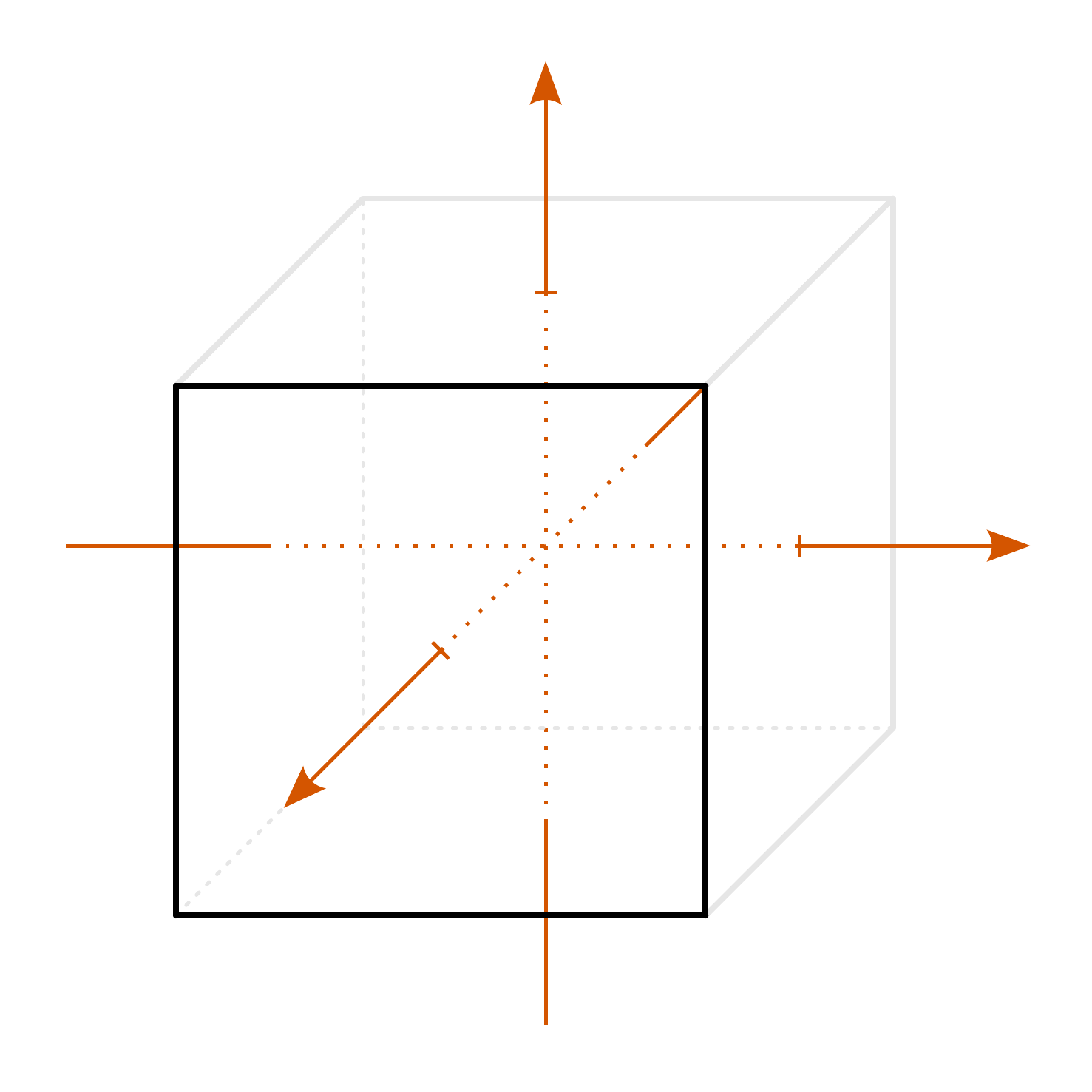}}%
    \put(0.52837723,0.72586047){\makebox(0,0)[lt]{\smash{\begin{tabular}[t]{l}$1$\end{tabular}}}}%
    \put(0.72723994,0.44442952){\makebox(0,0)[lt]{\smash{\begin{tabular}[t]{l}$1$\end{tabular}}}}%
    \put(0.37132678,0.42770549){\makebox(0,0)[lt]{\smash{\begin{tabular}[t]{l}$1$\end{tabular}}}}%
    \put(0,0){\includegraphics[width=\unitlength,page=2]{gswd_visual_5.pdf}}%
    \put(0.39784098,0.24305705){\makebox(0,0)[lt]{\smash{\begin{tabular}[t]{l}$\vec z_1$\end{tabular}}}}%
    \put(0.17926895,0.28269801){\makebox(0,0)[lt]{\smash{\begin{tabular}[t]{l}$\vec z_2$\end{tabular}}}}%
    \put(0,0){\includegraphics[width=\unitlength,page=3]{gswd_visual_5.pdf}}%
  \end{picture}%
\endgroup%

         \caption{}
     \end{subfigure}
     \hfill
     \begin{subfigure}[b]{0.3\textwidth}
         \centering
         \footnotesize
         \def\svgwidth{\textwidth}
\begingroup%
  \makeatletter%
  \providecommand\color[2][]{%
    \errmessage{(Inkscape) Color is used for the text in Inkscape, but the package 'color.sty' is not loaded}%
    \renewcommand\color[2][]{}%
  }%
  \providecommand\transparent[1]{%
    \errmessage{(Inkscape) Transparency is used (non-zero) for the text in Inkscape, but the package 'transparent.sty' is not loaded}%
    \renewcommand\transparent[1]{}%
  }%
  \providecommand\rotatebox[2]{#2}%
  \newcommand*\fsize{\dimexpr\f@size pt\relax}%
  \newcommand*\lineheight[1]{\fontsize{\fsize}{#1\fsize}\selectfont}%
  \ifx\svgwidth\undefined%
    \setlength{\unitlength}{425.19685039bp}%
    \ifx\svgscale\undefined%
      \relax%
    \else%
      \setlength{\unitlength}{\unitlength * \real{\svgscale}}%
    \fi%
  \else%
    \setlength{\unitlength}{\svgwidth}%
  \fi%
  \global\let\svgwidth\undefined%
  \global\let\svgscale\undefined%
  \makeatother%
  \begin{picture}(1,1)%
    \lineheight{1}%
    \setlength\tabcolsep{0pt}%
    \put(0,0){\includegraphics[width=\unitlength,page=1]{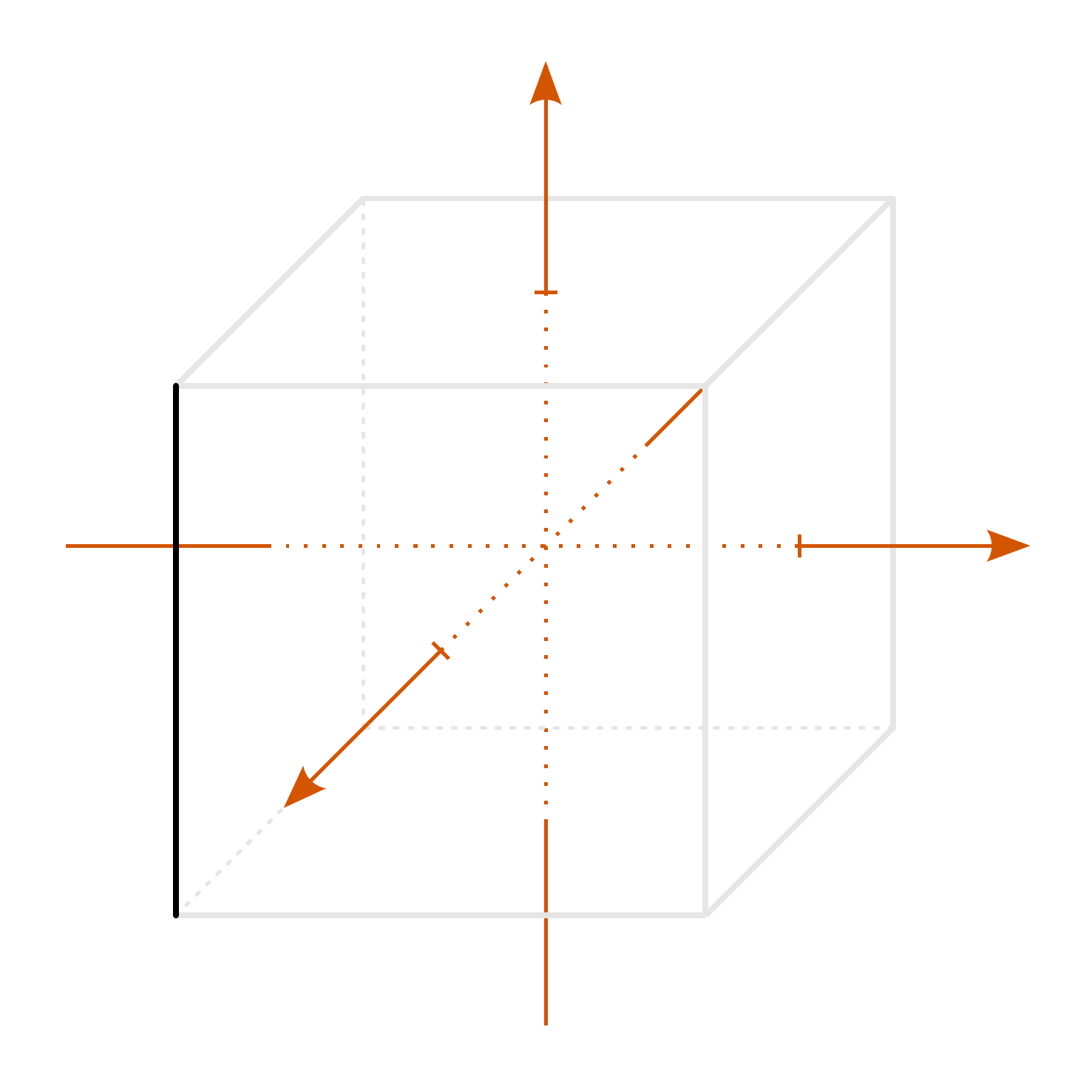}}%
    \put(0.52837723,0.72586047){\makebox(0,0)[lt]{\smash{\begin{tabular}[t]{l}$1$\end{tabular}}}}%
    \put(0.72723994,0.44442952){\makebox(0,0)[lt]{\smash{\begin{tabular}[t]{l}$1$\end{tabular}}}}%
    \put(0.4273259,0.37468541){\makebox(0,0)[lt]{\smash{\begin{tabular}[t]{l}$1$\end{tabular}}}}%
    \put(0,0){\includegraphics[width=\unitlength,page=2]{gswd_visual_6.pdf}}%
    \put(0.11383708,0.67024246){\makebox(0,0)[lt]{\smash{\begin{tabular}[t]{l}$\vec z_n$\end{tabular}}}}%
  \end{picture}%
\endgroup%

         \caption{}
     \end{subfigure} \\[1em]
     \caption[caption]{The procedure of the Gram-Schmidt Walk Design: The initial $\vec{z}_0 = \vec 0$ evolves step by step to a random $\vec{z}_t \in \{ \pm 1\}^n$, ensuring at the same time small covariate imbalance between the sets $Z^+$ and $Z^-$.\\\hspace{\textwidth}
        \textbf{(a)} Initially, $\vec z_0 = \vec 0$.
        \textbf{(b)} A direction $\vec u$ is chosen.
        \textbf{(c)} There are two step sizes to increase the integrality of $\vec z$ by walking along $\vec u$: Either in positive (green) or negative (red) direction, until a new boundary is hit. This step size is chosen at random.
        \textbf{(d)} In this case, the negative direction was chosen. $\vec z_1$ is "more integral" than $\vec z_0$.
        \textbf{(e)} We repeat this process.
        \textbf{(f)} After at most $n$ iterations, $\vec z$ is integral.}
        \label{fig:gswd-intuition}
       
\end{figure}

The Gram-Schmidt Walk Design starts by a relaxation: $\vec z  = \vec 0 \in [-1, +1]^n$. In order to achieve integrality, we first choose a direction $\vec u$. Then, we start walking from $\vec z$ randomly either in positive or negative direction along $\vec u$, until we hit a new boundary of $[-1, +1]^n$. By repeating this procedure at most $n$ times, we achieve integrality.
This process is depicted in \Cref{fig:gswd-intuition}.

In this procedure, we have the following two degrees of freedom that both help achieve a particular purpose:

\begin{itemize}
	\item The step direction $\vec u$ is chosen such that the covariate imbalance between $Z^+$, $Z^-$ does not increase by too much.\footnote{Note: In the beginning, $\vec z = \vec 0$, thus $Z^+ = Z^- = \varnothing$, and therefore the covariate imbalance between both sets is $0$. }
	\item The step size is chosen randomly between two values: Either positive or negative in order to hit a new boundary of $[-1, +1]^n$ when walking along $\vec u$. The probability distribution between these two step sizes is chosen such that the Horvitz-Thompson estimator is unbiased.
\end{itemize}

\subsection{The Algorithm} \label{sec:gswd-algo}
We restate the Gram-Schmidt Walk Design algorithm from Harshaw et al.\ in \Cref{alg:gswd}.

\begin{algorithm}
\caption{The Gram-Schmidt Walk Design (aka GSWD) from \cite{harshaw-BalancingCovariatesRandomized-2021}}

\label{alg:gswd}
    \algorithminput{The column vectors $\vec{b}_1, \ldots \vec{b}_n \; \in \R^{d+n}$ of $\mat B$ }
    
    \algorithmoutput{$z \in \{\pm1 \}^n$}
    
    \begin{algorithmic}[1]
    	\State{$t \leftarrow 0$}
    	\State{$\vec{z}_t \leftarrow (0, \ldots ,0)$}
    	\State{Select u.a.r. a unit $p \in [n]$ as the first \text{pivot} unit}
    	\While{$\vec{z}_t \notin \{\pm 1\}^n$}
    		\State{Create \emph{alive set} $\mathcal A \leftarrow \{ i \in [n]: |\vec{z}_t(i)|<1\}$}
    		\State{\textbf{if} $p \notin \mathcal A$ \textbf{then} select u.a.r. new $p \in \mathcal A$}
    		\State{Set \emph{step direction}
    		$$ \vec{u}_t \leftarrow \argmin_{\vec{u} \in U}{\norm{\mat B \vec u}}_2  $$
    		\qquad where
    		$$ U := \{\vec u \in \R^n\;\; \vert  \; \;\vec u(p) = 1 \; \land \; \forall i \notin \mathcal A: \vec u(i) = 0 \}.$$
    		 }
    		 \State{$\delta_t^+ \leftarrow |\max{\Delta_t}|, \; \delta_t^- \leftarrow |\min{\Delta_t}|$ where
    		 $$ \Delta_t := \{\delta \in \R \; \; \vert \; \; \vec{z}_t + \delta \vec{u}_t \in [\pm 1]^n\}$$
    		 }
    		 \State{Set \emph{step size} $\delta_t$:
    		 $$ \delta_t \leftarrow \begin{cases}
    		 	\delta_t^+ & \text{with probability }\frac{\delta_t^-}{\delta_t^+ + \delta_t^-}\\
    		 	-\delta_t^- & \text{with probability }\frac{\delta_t^+}{\delta_t^+ + \delta_t^-}
    		 \end{cases}$$
    		 }
    		 \State{Update fractional assignment $\vec{z}_{t+1} \leftarrow \vec{z}_t + \delta_t \vec{u}_t$}
    		 \State{$t\leftarrow t+1$}
    	\EndWhile{}
    	\State{\textbf{return} assignment vector $\vec{z}:=\vec{z}_t$}

    \end{algorithmic}
\end{algorithm}

Note that we did not emphasize in the intuitive explanation the procedure of choosing a pivot at random. While so-called "pivot phases" are important in the analysis of the Gram-Schmidt Walk in \cite{bansal-GramSchmidtWalkCure-2017}, the only importance of choosing these pivots \emph{at random} is for finding an estimator of the ridge loss, which is in turn necessary for constructing confidence intervals.\footnote{Formally, the random choice of pivots is used to prove that all "second-order assignment probabilities": $\forall i \neq j \in [n], v \in \{0,1\}^2: \Pr{(z_i, z_j) = v}$ are bounded away from zero. See Appendix A4.2 in \cite{harshaw-BalancingCovariatesRandomized-2021}} Therefore, we will not dive deeper into this.

A rigorous algorithm analysis can be found in the appendix of Harshaw et al. We highlight the key results in the following. 

\subsection{Analysis: Unbiasedness} \label{sec:gswd-analysis-bias}

\begin{theorem}\label{thm:gswd-ht-unbiased}
	Under the Gram-Schmidt Walk Design (\Cref{alg:gswd}), the Horvitz-Thompson estimator $\tauht$ is an unbiased estimator for the average treatment effect.
\end{theorem}

To prove \Cref{thm:gswd-ht-unbiased}, we first show the following lemma.

\begin{lemma}[Lemma 2 in \cite{harshaw-BalancingCovariatesRandomized-2021}] \label{lem:martingale-fractional-assignments}
	The sequence of fractional assignments in \Cref{alg:gswd} $\vec{z}_0, \vec{z}_1, \ldots$ forms a martingale.
\end{lemma}

\begin{proof}
The central observation is that by choice of $\delta_t$, we have
\begin{equation} \label{eq:step-size-in-exp-zero}
	\E{\delta_t \vert \delta_t^+, \delta_t^-} = \delta_t^+ \frac{\delta_t^-}{\delta_t^+ + \delta_t^-} - \delta_t^- \frac{\delta_t^+}{\delta_t^+ + \delta_t^-} = 0. 
\end{equation}

The rest follows by applying basic probabilistic identities, as done in the following.
\begin{align}
	\E{\vec{z}_{t+1} \vert \vec{z}_0,\ldots,\vec{z}_t} 
	&= \vec{z}_t + \E{\delta_t \vec{u}_t \vert  \vec{z}_0,\ldots,\vec{z}_t} \tag{linearity of $\mathbb{E}$} \\
	&= \vec{z}_t + \E{\E{\delta_t \vec{u}_t \vert \delta_t^+, \delta_t^-} \vert  \vec{z}_0,\ldots,\vec{z}_t} \tag{by the law of total $\mathbb{E}$}\\
	&= \vec{z}_t + \E{\E{\delta_t \vert \delta_t^+, \delta_t^-} \vec{u}_t \vert  \vec{z}_0,\ldots,\vec{z}_t} \tag{by conditional independence} \\
	&= \vec{z}_t \tag{applying Equation \ref{eq:step-size-in-exp-zero}}
\end{align}

where the conditional independence holds because $\delta_t$ is complety determined by $\delta_t^+, \delta_t^-$ and thus 
\begin{align}
	\Pr{ \delta_t = x \vert \vec{u}_t, \delta_t^+, \delta_t^-, \vec{z}_0, \ldots , \vec{z}_t} = \Pr{ \delta_t =x \vert \delta_t^+, \delta_t^-, \vec{z}_0, \ldots , \vec{z}_t}
\end{align}
for any $x \in \R$.
	
\end{proof}

\begin{proof}[Proof of \Cref{thm:gswd-ht-unbiased}] The unbiasedness of $\tauht$ under the Gram-Schmidt Walk Design follows now from the fact that $\vec{z}_1, \vec{z}_2,\ldots$ form a martingale and by observing that $\vec{z}_0 = \vec 0$. For the advanced reader familiar with basic martingale properties, we recommend skipping the remainder of this proof. 

	Formally, we show by induction that for any $t \geq 0$ (where $t$ is the index of an iteration of the GSWD):
	\begin{equation} \label{eq:martingale-induction}
		\E{\vec{z}_{t} \vert \vec{z}_0} = \E{\vec{z}_0} .
	\end{equation}
	
	Base case ($t=0$).
	\begin{align}
		\E{\vec{z}_{0} \vert \vec{z}_0}
		&=\E{\vec{z}_{0}}.
	\end{align}
	
	Step ($t>0$). Assume, by induction, that Equation (\ref{eq:martingale-induction}) holds for $t$. Then, we have
	\begin{align*}
		\E{\vec{z}_{t+1} \vert \vec{z}_0} 
		&=\E{\E{\vec{z}_{t+1}\vert \vec{z}_0, \ldots, \vec{z}_t} \vert \vec{z}_0} \tag{law of total expectation} \\
		&=\E{\vec{z}_t \vert \vec{z}_0} \tag{martingale property, see \Cref{lem:martingale-fractional-assignments}} \\
		&= \E{\vec{z}_0}. \tag{induction hypothesis, Equation \ref{eq:martingale-induction}}
	\end{align*}
	
	Therefore, we have for the algorithm's returned assignment vector $\vec{z}:$
	\begin{align*}
		\E{\vec{z}} 
		&= \E{\vec{z}| \vec{z}_0 = \vec{0}} \tag{by \Cref{alg:gswd}} \\
		&= \vec{0} \tag{by Equation \ref{eq:martingale-induction}}
	\end{align*} 
	Therefore, for any unit $i \in [n]$:
	$$ 0 = \E{\vec{z}(i)} = 1 \cdot \Pr{\vec{z}(i) = + 1} - 1 \cdot \Pr{\vec{z}(i) = + 1}$$
	which implies that $\Pr{\vec{z}(i) = + 1} =\Pr{\vec{z}(i) = - 1} = \frac{1}{2}$.
	From \Cref{lem:tauht-unbiased} we know that this implies unbiasedness of $\tauht$.
\end{proof}

\subsection{Analysis: Worst-Case Variance} \label{sec:gswd-analysis-variance}

This section shows two bounds on the worst-case variance of $\tauht$ under the Gram-Schmidt Walk Design.

In the following, we assume $\Phi \in (0,1]$ to eliminate issues when $\Phi = 0$.

\begin{theorem}[Theorem 1 in \cite{harshaw-BalancingCovariatesRandomized-2021}] 
\label{thm:covbz-and-projection-mtx}
Under the Gram-Schmidt Walk Design,
\begin{equation*}
	\Cov{\mat B \vec z} \preceq \mat P = \mat B (\mat B' \mat B)^{-1} \mat B'
\end{equation*}
\end{theorem}

\begin{proof}[Proof outline]
This proof is based on the main proof of Bansal et al.'s Gram-Schmidt Walk \cite{bansal-GramSchmidtWalkCure-2017} and stems from the connection of the above algorithm to its namesake, the Gram-Schmidt orthogonalization procedure. The lengthy proof is beyond the scope of this work and we refer to Appendix A, pp.\ 58-67, in \cite{harshaw-BalancingCovariatesRandomized-2021}. However, we give a basic sketch that should help the interested reader while studying the proof from Harshaw et al.

In general, the goal is to show a Loewner order bound, it thus suffices to show
\begin{equation}
	\forall v \in \R^{n+d}: \quad \vec v'\Cov{\mat B \vec z} \vec v \leq \vec v' \mat P \vec v.
\end{equation}
 
The key step is that for any $v \in \R^{n+d}$, $\vec v' \Cov{\mat B \vec z} \vec v$ can be rewritten in terms of the step sizes and directions, and then further in terms of projection matrices:

\newcommand{\rotleq}{\rotatebox[origin=c]{-90}{$\leq$}}

\begin{align*}
	\vec v' \Cov{\mat B \vec z} \vec v &= \sum_{i=1}^{n} \underbrace{\E{\sum_{t \in S_i} \delta_t^2 \langle \mat B \vec u_t, \vec v \rangle^2}}_{\leq \: \E{\vec v' \mat P_i \vec v}} \\
	&\leq \sum_{i=1}^n  \vec v' \E{\mat P_i }\vec v \\
	& =  \vec v'  \sum_{i=1}^n \E{\mat P_i }\vec v = \vec v'  \mat P \vec v.
\end{align*}
Here, $S_i$ is unit $i$'s pivot phase (the set of iterations for which unit $i$ was the pivot).\footnote{Note that $S_i$ can be empty or even contain multiple iterations. The latter can occur when some other unit's integrality constraint has been satisfied before unit $i$'s integrality constraint while walking along the step direction.} $\mat P_i$ is the projection matrix onto the subspace of $\left\{ \mat B \vec u_t: t \in S_i \right\}$, i.e., the subspace of updates generated during the $i$th pivot phase. The above equation shall serve as a guideline when studying the detailed proof of Harshaw et al., where proofs for all the steps are given.

To briefly touch on the connection to the Gram-Schmidt orthogonalization process, where a set of vectors are orthogonalized one after the other, it is shown in Bansal et al.\ that the subspaces mentioned above are orthogonal to each other, i.e.
\begin{equation}
	\operatorname{range}\left( \mat P_i \right) \perp \operatorname{range}\left(\mat P_j\right) \quad \quad \text{for }i \neq j \:\in [n] .
\end{equation}

\end{proof}

\begin{lemma} \label{lem:cov-z-bound-Q}
	\begin{equation*}
		\Cov{\vec z} \preceq \mat Q := (\Phi \mat I + (1- \Phi) \xi^{-2} \mat X \mat X' )^{-1}
	\end{equation*}
\end{lemma}

\begin{proof}
	We first calculate the upper $n \times n$ block of the projection matrix $\mat P$. The lemma then follows from this $n \times n$ block and \Cref{thm:covbz-and-projection-mtx}.
	
		By definition of the projection matrix $\mat P$ onto the column space of $\mat B$, we have
	\begin{align}
		\mat P 
		&= \mat B (\mat B' \mat B)^{-1} \mat B' \tag{def. of proj. matrix}\\
		&= \begin{bmatrix}
			\sqrt{\Phi} \mat I & \xi^{-1}\sqrt{1- \Phi}\mat X
		\end{bmatrix}
		\cdot
		\begin{bmatrix}
			\Phi \mat I + \xi^{-2} (1- \Phi)\mat X \mat X'
		\end{bmatrix}^{-1}
		\cdot
		\begin{bmatrix}
			\sqrt{\Phi} \mat I \\ \xi^{-1}\sqrt{1- \Phi}\mat X'
		\end{bmatrix} \tag{def. $\mat B$} 
	\end{align}
	
	and therefore, by matrix block multiplication (note that the middle matrix has dimension $n \times n$):
	\begin{equation}
		\mat P_{[1:n, 1:n]}=\Phi \left(\Phi \mat I + (1- \Phi) \xi^{-2} \mat X \mat X' \right)^{-1}.
	\end{equation}
	
	This inverse indeed exists, as for any $\Phi \in (0,1]$, $\Phi \mat I + (1- \Phi) \xi^{-2} \mat X \mat X'$ is positive definite and thereby invertible.
	
	By definition of $\mat B$, we have:
	\begin{equation}
		\Cov{\mat B \vec z} = \begin{bmatrix}
			\Phi \Cov{\vec z} 
			& \xi^{-1} \sqrt{\Phi (1-\Phi) \Cov{\mat X' \vec z, \vec z }'} \\ 
			\xi^{-1} \sqrt{\Phi (1-\Phi) \Cov{\mat X' \vec z, \vec z }} 
			& \xi^{-2} (1-\Phi) \Cov{\mat X' \vec z}
		\end{bmatrix} 
	\end{equation}
	
	As $\Cov{\mat B \vec z} \preceq \mat P $ (\Cref{thm:covbz-and-projection-mtx}), the same holds for the upper $n \times n$ block:
	\begin{equation}
		\Cov{\mat B \vec z}_{[1:n, 1:n]} \preceq \mat P_{[1:n, 1:n]}
	\end{equation}
	  (this follows by basic properties of the Loewner norm, for a formal proof see \Cref{lem:loewner-order-for-submatrices} in the Appendix), and thus 
	\begin{equation}
		\Phi \Cov{\vec z} \preceq \Phi (\Phi \mat I + (1- \Phi) \xi^{-2} \mat X \mat X' )^{-1}
	\end{equation}
	As $\Phi$ is positive, we can multiply by $\Phi^{-1}$, which gives the result.
\end{proof}

From this, the following (first) bound on the variance of $\tauht$ follows.

\begin{corollary}[Theorem 2 in \cite{harshaw-BalancingCovariatesRandomized-2021}] \label{cor:variance-bound-by-spectral-discussion}
	For any given trade-off parameter $\Phi \in (0,1]$ and potential outcome vector $\vec \mu = \frac{\vec a+\vec b}{2}$, the worst-case variance under the Gram-Schmidt Walk Design is upper bounded by:
	
	\begin{equation*}
		\Var{\tauht} \leq \frac{4}{\Phi n^2} \norm{\vec \mu}_2^2
	\end{equation*}
\end{corollary}

\begin{proof} This result follows from a bound in the spectral discussion and a bound on the maximal eigenvalue.

We know from the spectral discussion  (\Cref{lem:worst-case-variance-of-a-design}), that 
\begin{equation} \label{eq:detail:variance-bound-spectral-recall}
	   \Var{\tauht} \leq \frac{4 \norm{\vec \mu}_2^2}{n^2} \lambda_{max}
\end{equation}
where $\lambda_{max}$ is the maximal eigenvector of $\Cov{\vec z}$. As $\Cov{\vec z} \preceq \left(\Phi \mat I + (1- \Phi) \xi^{-2} \mat X \mat X' \right)^{-1}$ (\Cref{lem:cov-z-bound-Q}), we have
\begin{align*}
	\lambda_{max}^{-1}\left(\Cov{\vec z} \right)
	&\geq \lambda_{max}^{-1} \left( \left(  \Phi \mat I + (1- \Phi) \xi^{-2} \mat X \mat X'\right)^{-1} \right) \tag{due to Loewner order} \\
	&= \lambda_{min}\left( \Phi \mat I + (1- \Phi) \xi^{-2} \mat X \mat X'\right) \\
	&= \Phi + (1-\Phi) \xi^{-2} \lambda_{min}\left(\mat X \mat X'\right) \\
	&\geq \Phi \tag{as $\mat X \mat X'$ is PSD and $(1-\Phi) \xi^{-2} \geq 0$}
	.
\end{align*}

Therefore,
\begin{equation} \label{eq:detail:lambda-max-bound}
	\lambda_{max}\left(\Cov{\vec z} \right)  \leq \frac{1}{\Phi} 
\end{equation}
and we can bound the variance
\begin{equation}
	 \Var{\tauht} \stackrel{(\ref{eq:detail:variance-bound-spectral-recall})}{\leq} \frac{4}{n^2}  \norm{\vec \mu}_2^2 \lambda_{max}  \stackrel{(\ref{eq:detail:lambda-max-bound})}{\leq} \frac{4}{\Phi n^2} \norm{\vec \mu}_2^2.
\end{equation}
\end{proof}

We can summarize the contributions that lead to \Cref{cor:variance-bound-by-spectral-discussion} in the following diagram.

\begin{figure}[H]
\centering
\begin{equation*}
\begin{tikzcd}[column sep=-2pt,
every arrow/.append style={dash}]
	& & \mbox{} \arrow[d,rightarrow,"\text{\Cref{thm:covbz-and-projection-mtx}}" description] \\
	& \Cov{\mat B \vec z} & \preceq \arrow[d,rightarrow,"\text{aka}" description]& \mat P \\
	& \begin{pmatrix}
		\Phi \cdot \Cov{\vec z} & \\
		& \\
	\end{pmatrix} & \preceq \arrow[d,Rightarrow] & \begin{pmatrix}
		\Phi \cdot \mat Q & \\
		& \\
	\end{pmatrix}  \\
	& \lambda_{max} \left( \Cov{\vec z} \right) & \leq \arrow[d,Rightarrow] & \lambda_{max} \left(\mat Q \right) \\
	\Var{\taunet} \stackrel{\Cref{lem:worst-case-variance-of-a-design}}{\leq} & \frac{4 \norm{\vec \mu}_2^2}{n^2}\lambda_{max} \left( \Cov{\vec z} \right) & \leq & \frac{4 \norm{\vec \mu}_2^2}{n^2}\lambda_{max} \left( \mat Q \right)
\end{tikzcd}
\end{equation*}
\caption{The proof of \Cref{cor:variance-bound-by-spectral-discussion}.}
\end{figure}

However, it is possible to refine this result. In order to do so, we first need to establish a connection to the ridge regression loss.

\begin{lemma}[Lemma A10 in \cite{harshaw-BalancingCovariatesRandomized-2021}]
\label{lem:connection-gswd-ridge}
For any $\vec \mu \in \R^n$, $\Phi \in (0,1]$ and $\mat X  \in \R^{n,d}$ with maximum row norm $\xi = \max_{i \in [n]}\norm{\vec x_i}_2$, the following equality holds.
\begin{equation}
	\vec \mu' \mat Q \vec \mu = \vec \mu' (\Phi \mat I + (1- \Phi) \xi^{-2} \mat X \mat X' )^{-1} \vec \mu = \min_{\vec \beta \in \R^d}\left( \frac{1}{\Phi} \norm{\vec \mu - \mat X \vec \beta}^2 + \frac{\xi^2}{1-\Phi} \norm{\vec \beta}^2 \right)
\end{equation}

\end{lemma}

\begin{proof}
	Note that there exists an explicit formula for the optimal $\vec \beta \in \R^d$, see Hastie et al.\ \cite{hastie-ElementsStatisticalLearning-2009}. This lemma follows then by realizing that the optimal value of the ridge loss function, i.e., the optimizer $\vec \beta \in \R^d$ plugged into the ridge loss function, is equal to $\vec \mu' \mat Q \vec \mu$. 
	
	For a detailed multi-page calculation, we refer to pp. 79-81 in the appendix of \cite{harshaw-BalancingCovariatesRandomized-2021}.
\end{proof}

This lemma allows now for an improvement of \Cref{cor:variance-bound-by-spectral-discussion}:

\begin{theorem}[Theorem 3 in \cite{harshaw-BalancingCovariatesRandomized-2021}] \label{thm:variance-bound-gswd-ridge}
	For any given trade-off parameter $\Phi \in (0,1)$ and potential outcome vector $\vec \mu = \frac{\vec a+\vec b}{2}$, the worst-case variance under the Gram-Schmidt Walk Design is upper bounded by:
	\begin{equation*}
		\Var{\tauht} \leq \frac{4}{n^2} \min_{\vec \beta \in \R^d}\left( \frac{1}{\Phi} \norm{\vec \mu - \mat X \vec \beta}^2_2 + \frac{\xi^2}{1-\Phi} \norm{\vec \beta}^2_2 \right)
	\end{equation*}
\end{theorem}

\begin{proof}
This result follows by first rewriting the variance in terms of the squared error expression (\Cref{lem:error-of-tau-ht}) and then applying the connection to the ridge loss established in \Cref{lem:connection-gswd-ridge}:
	\begin{align*}
		\frac{n^2}{4} \Var{\tauht} 
		&\stackrel{\Cref{lem:error-of-tau-ht}}{=} \vec \mu' \Cov{\vec z} \vec \mu \\ 
		&\stackrel{\Cref{lem:cov-z-bound-Q}}{\leq}  \vec \mu' \mat Q \vec \mu 
		\stackrel{\Cref{lem:connection-gswd-ridge}}{=} \min_{\vec \beta \in \R^d}\left( \frac{1}{\Phi} \norm{\vec \mu - \mat X \vec \beta}^2_2 + \frac{\xi^2}{1-\Phi} \norm{\vec \beta}^2_2 \right).
	\end{align*}
\end{proof}

\Cref{thm:variance-bound-gswd-ridge} lets us connect to previous results and an earlier discussion from the spectral section. First, notice that \Cref{thm:variance-bound-gswd-ridge} is at least as good as \Cref{cor:variance-bound-by-spectral-discussion}: Choosing $\vec \beta  = \vec 0$ makes their bounds match. 

But \Cref{thm:variance-bound-gswd-ridge}'s power can be discovered when considering $\Phi \rightarrow 0$. Recall that this choice of $\Phi$ corresponds to emphasizing ``potential performance'' in the potential performance versus robustness tradeoff. Then, the term
\begin{equation*}
	\frac{1}{\Phi} \norm{\vec \mu - \mat X \vec \beta}^2_2
\end{equation*}
has the most weight in \Cref{thm:variance-bound-gswd-ridge}. If the covariates are linearly predictive of the potential outcomes vector, we have that $\vec \mu$ is close to the column span of $\mat X$, and we can thus find a good $\vec \beta$ such that $ \norm{\vec \mu - \mat X \vec \beta}^2_2 \approx 0$. Thereby, \Cref{thm:variance-bound-gswd-ridge} gives a good upper bound for the variance of $\tauht$ -- which we have shown to be the mean squared error of the estimator. 

Because we know that the variance can be decreased with an increase of the experiment size $n$, this result is significant when increasing the sample size of an experiment is expensive or not possible.

\newpage

\chapter{Network Effects} \label{sec:network-effects}

A key assumption in many RCTs, including the model considered in Harshaw et al., relies on Rubin's SUTVA (see \Cref{sec:SUTVA}): The absence of any influence between units. However, in many situations, this is an inaccurate model that can lead to wrong results. For example, in a vaccine trial, a participant is not only influenced by his own group assignment (vaccine / placebo) but also by the vaccination rate in his peer group. Therefore, the underlying social network should be considered in an effective RCT.

\section{Literature and Overview}

The influence of one unit on another unit's observed outcome is commonly referred to as \textdef{interference} or \textdef{spillover effect} (the effect on one individual ``spills over'' to another) and occurs in literature in two ways: It might either be \emph{the quantity of interest} in an experiment (how do peer groups affect each other?) or an undesired perturbation factor that makes the statistical analysis of an RCT harder. The latter is the case for the vaccine trial described in the beginning, and is the focus of this work.

Early occurrences of interference are in agricultural studies, where nearby crops influenced each other. This carried over to educational research, where disjoint student groups are exposed over an extended period of time to different learning techniques, but any interference (e.g., by communication between students from different groups) should be ruled out. A common practice in these educational studies is block randomization at the institutional level, where schools are treated as blocks, and each school is randomly assigned to treatment / control \emph{as a whole}\footnote{Note that this is different from the ``Permuted-Block Randomization'' scheme described in \Cref{sec:permuted-block-randomization}, where the goal was to get balanced assignments \emph{within} each block.} \cite{schochet-WhatDesignBasedCausal-2017}.
 The assumption made there is that students might communicate within schools, but there is no communication between students from different schools. Sobel calls this ``partial interference assumption'' \cite{sobel-WhatRandomizedStudies-2006}, which has been studied extensively \cite{hudgens-CausalInferenceInterference-2008}. But also this model's field of application is limited. Coming back to the vaccine trial, what are the blocks? Contagious diseases spread via social contacts within a population. Thus, there is a social network underlying the spillover effects, and we therefore speak of \textdef{network interference}. As we model \emph{potential} influence (it is not known ex-ante which person will infect another), we have a rather \emph{connected} graph instead of multiple, clearly separated clusters. The example of a vaccine trial motivates the need for both an RCT model as well as a treatment effect estimator under network interference. 
 
 Before designing an appropriate estimator, we need to model how units influence each other. In a recent economics paper, Candogan et al.\ model interference patterns as a graph and consider interference as soon as a unit $u$ has any neighbor not in the same treatment as $u$. If this is the case, they discard the data obtained from $u$. If this is not the case ($u$ has only neighbors within the same treatment group), they keep the observed outcome from $u$ as it is. \cite{candogan-NearOptimalExperimentalDesign-2021} 
 
 This model has three major drawbacks.
 Firstly, their unsustainable dealings with data points leads to low precision: A unit is likely to have at least one neighbor that is not in the same treatment group. By discarding the observed outcomes from any such units, the number of units used for analysis (call this $n'$) shrinks significantly, and so does the estimator's precision.\footnote{Intuitively, a small sample size $n'$ results in an imprecise estimation}
 
 Second, the intensity of network effects is not accounted for. Suppose there are two units (that have both only neighbors in the same treatment group and are therefore not discarded) where one of them has only a single neighbor, and the other has 100 neighbors. The latter might be more influenced than the former, but this is not accounted for in this model.
 
 Third, what if we only know the likelihood of interaction between two units? Their model assumes a deterministic graph, which might not be known ex-ante -- i.e., when making the group assignments.
 
The three points above show that it is crucial to define the model of interference thoughtfully. 

Further, Candogan et al.\ describe the challenge of finding an estimator under network interference as follows:

\begin{displayquote}
``In presence of network interference, even designing an unbiased estimator for $\tau$ given a randomized experiment can be challenging'' -- \cite{candogan-NearOptimalExperimentalDesign-2021}
\end{displayquote}

We overcome all of the issues described above in the following sections: First, we find a good interference model (\Cref{sec:modelling-network-spillover}) -- which is different from Candogan et al.\ Then, we present an unbiased estimator,\footnote{The estimator exists under certain realistic assumptions, for which we will give sufficient conditions} $\taunet$, for the average treatment effect (\Cref{sec:finding-taunet}). Next, we analyze its variance (\Cref{sec:taunet-error-analysis}), where we can give a nice matrix-vector product expression that bounds the variance (\Cref{thm:variance-under-bounds-assumption}). Then, we draw comparisons to Candogan et al.\ by giving a similar linear program as \cite{candogan-NearOptimalExperimentalDesign-2021} (\Cref{sec:worst-case-optimization-lp}) and finally discuss the role of disconnected graphs (\Cref{sec:disconnected-graph-block-design}).

\section{Modeling Network Spillover Effects}\label{sec:modelling-network-spillover}

Finding a good model is the foundational part of this work: Without a good model, the subsequent analysis is useless. The process of finding a good model is a balancing act between generality and complexity: The good model should be both general enough to be representative of the real world and simple enough to be analyzable. 

We outline in this section different possible models and give thereby reasons for our final model.

\subsection{Finding the Basic Model}
Suppose we knew the underlying social network in the form of a graph $G=(V,E)$ with $V = [n]$ and adjacency matrix $\mat A$. Without loss of generality, we assume no self-loops.\footnote{A self-loop at node $i$ would mean that unit $i$ influences itself, which is simply this unit's base outcome, $y_i$}

Under interference, we do not observe only this unit's \emph{base outcome} $y_i$ but also some network effect based on its neighbors. In our model, we assume that these parts are additive,\footnote{The assumption of additivity is mainly based on the goal to have a model that is simple enough to be analyzed: An additive network effect both allows us to use linearity of expectation and rewrite it in terms of a matrix-vector product.} so that we get for the \emph{observed outcome} under interference, $y_i'$:
\begin{equation} \label{eq:basic-def-of-yprime}
	\underbrace{y'_i}_{\substack{\text{observed outcome,}\\ \text{under interference}}} = \underbrace{y_i}_{\text{base outcome}} +  \underbrace{\text{some quantity depending on unit $i$'s neighbors and $\mat A$}}_{\text{network effect}}
\end{equation}

The question at hand is: What is a good quantity for ``some quantity depending on unit $i$'s neighbors and $\mat A$''?
We explore some possibilities in the following.

\paragraph*{Echo chamber model.}
\begin{equation}
	y_i' = y_i + z_i \sum_{j \in [n]} A_{i,j} z_j
\end{equation}

The idea of this model is similar to an echo chamber. One can think of people's political points of view and how they influence each other through communication. Consider, for example, a group of people belonging to either of two political parties, say party ``D'' and party ``R''. It is a well-known phenomenon that people that only communicate within their party (sociologically speaking, their ``bubble'') and never consider outside opinions, tend to have a more extreme point of view. On the other hand, if someone communicates with both ``R'' and ``D'' people, they might have a more balanced opinion, and their point of view does not get much distorted from their friends' opinion.

This model tries to capture exactly this notion. To see that quantitatively, consider a unit $i$ in group $Z^+$. If this unit has many neighbors within the same group, $Z^+$, the network effect term $+z_i \sum_{j \in [n]} A_{i,j} z_j$ is big and thus the observed outcome $y_i'$ increases.
On the other hand, if a unit $k$ in group $Z^-$ has many neighbors in $Z^-$, the term $+z_k \sum_{j \in [n]} A_{k,j} z_j$ is big (as $z_k = -1$) and thus the observed outcome $y_j'$ increases, too. For a unit $l$ with connections to both groups equally, the network term will be zero, and we observe $y_l' = y_l$.

This model has a specific statistical property:
\begin{equation}
	\E{\vec z}{z_i \sum_{j \in [n]} A_{i,j} z_j} = 0
\end{equation}
(using the fact that $G$ has no self-loops) and therefore, in expectation, $\vec y' = \vec y$.

However, this model has two major shortcomings. First, a unit's influence on others might depend on the magnitude of its base outcome $y_i$: People with a stronger opinion might have a stronger influence on their peer group. This model does only account for edge weights but not for the neighbor's base outcome. Accounting for such a dependence (adding a factor $y_j$ in the sum) would destroy the nice statistical property described above. 
Second, while this echo chamber model might seem reasonable in some sociological context, there are many other cases where network effect is inherently different. Recall that in this model, the observed outcome's increase depends not on the neighbors' \emph{actual} treatment groups but only on their group assignments \emph{relative} to the unit. It could also be that the neighbors' \emph{actual} treatment groups determine the influence on the observed outcome. For example, units from $Z^+$ might universally increase their neighbor's observed outcome -- independent of their group -- and units from $Z^-$ might universally decrease their neighbor's observed outcome.

\paragraph*{Treatment group dependent model.}
Fixing the above model's drawbacks, we consider the following model
\begin{equation}
	y_i' = y_i + \sum_{j \in [n]} A_{i,j} y_j.
\end{equation}

Here, the network influence also depends on the neighbor's base outcome ($y_j$) and it depends on the neighbor's \emph{actual} group assignment (versus in the echo chamber model, where the sum's $z_i$ coefficient made the effect depend only on the neighbor's group \emph{relative} to unit $i$). In vector notation, this model is 
\begin{equation}
	\vec y' = \vec \left(\mat I + \mat A \right)\vec y.
\end{equation}

Mathematically, we could simply infer $\vec y$ from $\vec y'$ (assuming $(\mat I+\mat A)$ is invertible): 
\begin{equation}
	\vec y = (\mat I+\mat A)^{-1}\vec y'
\end{equation}
and use the standard Horvitz-Thompson estimator on $\vec y$. Is modeling interference really that simple? Taking a close look shows us that it is not:

\paragraph*{Introducing randomness.}

Knowing connections between units (such as friendships or social closeness) does not mean that they really do influence each other. It is just a possibility. Nonetheless, we can say that the likelihood of influence is stronger for persons with a strong friendship or more communication. Therefore, the actual observed outcome under interference is rather a random variable, following some probabilistic model that we make of the network.

 To account for this, we model each edge weight as an independent random variable $C_{i,j} \geq 0$ with known expected value $A_{i,j}$. We do not make any further assumptions. We let this random model deliberately be that free and do not make any further assumptions on the exact distribution, as this is situation dependent.
 
 Our model now becomes:

\begin{equation}
	\vec{y'} = (\mat I + \mat{C}) \vec{y} \qquad \quad \text{with } \E{\mat C} = \mat A \;,\; \operatorname{diag}(\mat C) = \vec 0
\end{equation}

The experimenter will give the underlying expected adjacency matrix $\mat A$ and will define the probability distribution of $\mat C$. Note that $\operatorname{diag}(\mat C) = \vec 0$ corresponds to the absence of self-loops.

\subsection{Our Final Model}
Therefore, our final model becomes:
\begin{definition}[Probabilistic interference model]
	The observed outcome under interference is 
	$$ \vec{y'} = (\mat I + \mat{C}) \vec{y}$$
	where $\mat C$ is the random adjacency matrix with known expectation $\E{\mat C} = \mat A$, $\operatorname{diag}(\mat C) = \vec 0$ and $\vec y \in \R^n$ with
	\begin{equation*}
		 y_i = \begin{cases}
			a_i & \text{if } z_i = +1 \\
			b_i & \text{if } z_i = -1.
		\end{cases}
	\end{equation*}
\end{definition}

\subsection{Remarks on the model}

It shall be noted that another possibility would be to use 

\begin{equation}
	y_i' = y_i + \sum_{j \in [n]} A_{i,j} y_j' \quad \quad \text{(note the ``prime'' symbol: $y_j'$)}.
\end{equation}
which would reflect the inherent recursive nature of peer influence: One unit influences another, which again, under this influence, influences the next one.  There are two points to make on this idea. First, order-$n$ influence can be described already with the adjacency matrix $A$, as we do not restrict it to be binary. Therefore, further degree-$k$-neighborhoods in $G$ could be encoded using the $k$th power of the adjacency matrix instead. However, and this is the second point, this would create difficulties when adding randomness at a later stage. We discuss this further in \Cref{sec:outlook}.

\section{Our Average Treatment Effect Estimator} \label{sec:finding-taunet}
The Horvitz-Thompson estimator (\Cref{def:tauht}) simply applied to the observed outcomes $\vec y'$ would be biased, as it still contains all the spillover effects. Therefore, we define the following estimator for the (unobservable) average treatment effect:
\begin{definition}[Network ATE estimator]
	$$ \taunet := \frac{2}{n} \biggl< \vec z, \: (\mat I + \mat A)^{-1} \vec y' \biggr>$$
\end{definition}
 
Note that all quantities in this definition are known ($\mat A$, $n$) or measurable ($\vec y'$).

\subsection*{Well-definedness: Conditions on invertibility}
Our estimator has the nice property of being unbiased (see next section). Its well-definedness, however, depends on the invertibility of $\mat I + \mat{A}$. In the following, we give some sufficient conditions on the probabilistic graph model under which $\mat I + \mat{A}$ is guaranteed to be invertible. This should illustrate that under reasonable assumptions, $\tauht$ is well-defined.

Note that the exact modeling of the probabilistic graph will be the experimenter's task. This section aims to give examples of sufficient conditions on the graph model but does by no means try to capture all necessary conditions.

In general, it is useful to have some mechanism in the model that allows an edge $(i,j)$ to have zero weight with certainty, as big social networks inherently have clusters and thereby units that are known to have no interaction. The Bernoulli model provides that functionality.

\paragraph*{Bernoulli model.}

Suppose unit $j$'s influence on unit $i$ ($i \neq j$) is either $0$ or $\alpha$, according to a Bernoulli distribution:
\begin{equation}
	C_{i,j} = \begin{cases}
		\alpha & \text{with probability }p_{i,j} \\
		0 & \text{with probability }1-p_{i,j}. \\
	\end{cases}
\end{equation}

\begin{lemma}
	Under the Bernoulli model, $\taunet$ is guaranteed to be well-defined for
	\begin{equation*}
		\alpha  < \frac{1}{d_{\text{max}}}
	\end{equation*}
	where $d_\text{max}$ is the maximum degree that any node has in the graph with positive probability.
\end{lemma}

\begin{proof}
	Suppose $\alpha  < \frac{1}{d_{\text{max}}}$ where $d_\text{max}$ is the maximum degree that any node has with positive probability, i.e.
	\begin{equation} \label{eq:def-d-max}
		d_\text{max} := \max_{i \in {n}} \sum_{\substack{j\in[n]\\ j \neq i}} {\mathbbm{1}_{\{p_{i,j}>0\}}}.
	\end{equation}
	This gives for any $i \in [n]$:
	\begin{align}
		\sum_{\substack{j \in [n] \\ j \neq i}} A_{i,j} 
		&= \sum_{\substack{j \in [n] \\ j \neq i}} p_{i,j} \alpha \\
		&\leq d_\text{max} \alpha  \tag{by def. of $d_\text{max}$}\\
		&< 1 \tag{by condition on $\alpha$}.
	\end{align}
	Therefore, $\mat I + \mat A$ is strictly diagonally dominant, thereby $\left(\mat I + \mat A \right)^{-1}$ exists and thus $\taunet$ is well-defined.
\end{proof}

\paragraph*{Uniform weight distribution with activation.}

We can extend the Bernoulli model where not only the existence of a link is probabilistic, but also the weight -- if it exists. The following captures this in case of a uniform weight distribution
\begin{equation}
	C_{i,j} = \begin{cases}
		X_{i,j} & \text{with probability }p_{i,j} \\
		0 & \text{with probability }1-p_{i,j}\\
	\end{cases}
	\quad \text{where }X_{i,j} \sim \mathcal{U}[0, \alpha_{i,j}].
\end{equation}

Under this model, an edge $(i,j)$ exists with probability $p_{i,j}$. If it exists, its weight is uniformly distributed in $[0, \alpha_{i,j}]$.

\begin{lemma}
	Under the uniform weight distribution with activation model, $\taunet$ is guaranteed to be well-defined for
	\begin{equation*}
		\max_{i,j \in [n]}\alpha_{i,j}  < \frac{2}{d_{\text{max}}}
	\end{equation*}
	where $d_\text{max}$ is the maximum degree that any node has in the graph with positive probability.
\end{lemma}

\begin{proof}
	Suppose $\max_{i,j \in [n]}\alpha_{i,j} < \frac{2}{d_{\text{max}}}$ where $d_\text{max}$ is defined as in Equation \ref{eq:def-d-max}. 
		This gives for any $i \in [n]$:
	\begin{align}
		\sum_{\substack{j \in [n] \\ j \neq i}} A_{i,j} 
		&= \sum_{\substack{j \in [n] \\ j \neq i}} p_{i,j} \frac{\alpha_{i,j}}{2} \\
		&\leq d_\text{max} \max_{\substack{j \in [n] \\ j \neq i}} \frac{ \alpha_{i,j}}{2}  \tag{by def. of $d_\text{max}$}\\
		&< 1 \tag{by condition on $\max_{i,j \in [n]}\alpha_{i,j}$}.
	\end{align}
	Therefore, $\mat I + \mat A$ is strictly diagonally dominant, thereby $\left(\mat I + \mat A \right)^{-1}$ exists and thus $\taunet$ is well-defined.
\end{proof}

For a particular probabilistic graph model in an experiment, the experiment will either have to verify beforehand that under his assumptions, $\mat I + \mat A$ is guaranteed to be invertible, or they will have to check after determining the expected edge weights, denoted in $\mat A$, that the matrix $\mat I + \mat A$ is invertible.

From now on, we will assume well-definedness of $\tauht$.

\section{Error Analysis} \label{sec:taunet-error-analysis}

To analyze bias and variance of $\taunet$, we first need to find a compact expression of the estimator's error. To that end, we use the analysis of $\tau$ from Harshaw et al.\ (see proof of Lemma A1 in \cite{harshaw-BalancingCovariatesRandomized-2021}):

\begin{definition}
$\forall i \in [n]:$
	\begin {align*}
	\mathring a_i := \begin{cases}
		a_i & \text{if } z_i = +1 \\
		0 & \text{if } z_i = -1
	\end{cases}
	&&
	\mathring b_i := \begin{cases}
		0 & \text{if } z_i = +1 \\
		b_i & \text{if } z_i = -1
	\end{cases}
	\end{align*}
\end{definition}

It can be easily seen that the following identities hold.
\begin{lemma}[Identities for $\vec{\mathring a},\vec{\mathring b}$ as in \cite{harshaw-BalancingCovariatesRandomized-2021}] \label{lem:identities-ring-a-b} ~ 
\begin{itemize}
		\item $\vec y = \mathring{\vec a} + \mathring{\vec b}$ (by definition)
		\item $\langle \vec 1, \mathring{\vec a} \rangle = \langle \vec z,  \mathring {\vec a} \rangle$
		\item $\langle \vec 1, \mathring{\vec b} \rangle = \langle -\vec z, \mathring {\vec b} \rangle$
		\item $\langle \vec 1, \vec a - \mathring{\vec a} \rangle = \langle -\vec z, \vec a - \mathring {\vec a} \rangle$
		\item $\langle \vec 1, \vec b - \mathring{\vec b} \rangle = \langle -\vec z, \vec b - \mathring {\vec b} \rangle$
	\end{itemize}
\end{lemma}
Their purpose is to express the deterministic $\tau$ in terms of $\vec z$ to make it better comparable with $\taunet$. Therefore, we can express the average treatment effect as

\begin{align*}
	n \cdot \tau 
	&= \langle 1 , \vec a \rangle -  \langle 1 , \vec b \rangle  \tag{by definition} \\
	&= \langle 1 , \mathring{\vec a} + \vec a - \mathring{\vec a}\rangle -  \langle 1 ,\mathring{\vec b} + \vec b - \mathring{\vec b} \rangle \\
	&= \langle 1 , \mathring{\vec a} \rangle + \langle 1 , \vec a- \mathring{\vec a} \rangle -  \langle 1 ,\mathring{\vec b} \rangle -  \langle 1 ,\vec b - \mathring{\vec b} \rangle \\
	&= \langle \vec z , \mathring{\vec a} \rangle - \langle \vec z , \vec a- \mathring{\vec a} \rangle +  \langle \vec z ,\mathring{\vec b} \rangle -  \langle \vec z ,\vec b - \mathring{\vec b} \rangle \tag{by \Cref{lem:identities-ring-a-b}} \\
	&= \langle \vec z, 2 \mathring{\vec a} -\vec a + 2 \mathring{\vec b} - \vec b \rangle \\
	&= 2 \langle \vec z, \vec y \rangle - \langle \vec z, \vec a + \vec b \rangle. \tag{by \Cref{lem:identities-ring-a-b}}
\end{align*}

By defining $\mat E := (\mat I + \mat A)^{-1} (\mat I + \mat C) -\mat I$, our estimator $\taunet$ can be rewritten as follows.
\begin{align*}
	n \cdot \taunet 
	&= 2 \langle \vec z, (\mat I + \mat A)^{-1} \vec y' \rangle \tag{by definition of $\taunet$} \\
	&=  2 \langle \vec z, (\mat I + \mat A)^{-1}  (\mat I + \mat C) \vec y \rangle \tag{by definition of $\vec y'$}\\
	&= 2 \langle \vec z, \mat E \vec y \rangle + 2 \langle  \vec z, \vec y \rangle  \tag{by definition of $\mat E$}\\
\end{align*}

Therefore, the error $\taunet -\tau$ is
\begin{equation} \label{eq:taunet-error}
	\taunet -\tau = \frac{1}{n} \Big( \langle \vec z, \vec a + \vec b \rangle  + 2 \langle \vec z, \mat E \vec y \rangle \Big).
\end{equation}

\begin{remark}
Note that the only difference to the error of $\tauht$ in the non-interference case
\begin{equation*}
	\tauht - \tau = \frac{1}{n} \Big(  \langle \vec z, \vec a + \vec b \rangle  \Big) \tag{error in non-interference case}
\end{equation*}
 is the term $ +2 \langle \vec z, \mat E \vec y \rangle $.

 For $\tauht$, this error representation eliminates $\vec y$ in the right side of $\langle \vec z, \bullet \rangle $. This leads to having only linear dependencies on $\vec z$ in the $\langle \vec z, \bullet \rangle $ term. More importantly, it leads to having only quadratic terms in $\langle \vec z, \bullet \rangle^2 $, which is part of the variance expression.
 
 In our network interference case at hand, $\vec y$ does not cancel, so that the right-hand side of $\langle \vec z, \bullet \rangle $ depends on $\vec y$. As $\vec y$ depends on the random vector $\vec z$, we have quadratic dependence on $\vec z$ in $\langle \vec z, \bullet \rangle $ (relevant for the bias) and \emph{quartic} dependence in $\langle \vec z, \bullet \rangle^2 $ (relevant for the variance), making an analysis hard. However, we can eliminate the dependency in the bias term (\Cref{sec:taunet-unbiased})and reduce to \emph{cubic} dependencies in the variance term (\Cref{sec:taunet-variance}).
 \end{remark}
 
 \begin{remark}
 	The error of $\taunet$ (Equation \ref{eq:taunet-error}) depends on two inherently different random variables. On the one hand, it depends on the random matrix $\mat C$, on which $\mat E$ depends. On the other, it depends on the design $\vec z$ upon which $\vec y$ depends. But both $\mat C$ and $\vec z$ are \emph{stochastically} independent, as the former describes the random connections (e.g., friendships or communication), and the latter describes the group assignment: We can only make the design dependent on the observable $\mat A = \E{\mat C}$ but not on the unobservable $\mat C$. Therefore, $\mat C \perp \vec z$.
 \end{remark}
 
\subsection{Unbiasedness} \label{sec:taunet-unbiased}

\begin{theorem}\label{thm:taunet-unbiased}
	The average treatment effect estimator $\taunet$ is unbiased.
\end{theorem}

\begin{proof}
	\begin{align}
		n \cdot \E{\vec z, \mat C}{\taunet-\tau} 
		&= \E{\vec z}{\E{\mat C}{\langle \vec z, \vec a + \vec b \rangle  + 2 \langle \vec z, \mat E \vec y \rangle }} \tag{$\mat C \perp \vec z$} \\
		&= \E{\vec z}{\E{\mat C}{\langle \vec z, \vec a + \vec b \rangle}} \tag{$\E{\mat C}{\mat E} = 0$} \\
		&= \E{\vec z}{\langle \vec z, \vec a + \vec b \rangle} \tag{indep. of $\mat C$} \\
		&= 0 \tag{$\E{\vec z} = \vec0$}
	\end{align}
	
	Note that we used our general assumption, namely that the design ensures $\E{\vec z} = \vec0$.
\end{proof}

\subsection{Precision: The Estimator's Variance} \label{sec:taunet-variance}

\begin{lemma} \label{lem:taunet-variance-form-with-expectation}
	The variance of $\taunet$ is 
	\begin{equation*}
		\Var{\vec z, \mat C}{\taunet} = \frac{1}{n^2} \Big( (\vec a + \vec b)' \Cov{\vec z} (\vec a + \vec b) + 4 \E{\vec z, \mat C}{\left( \vec z' \mat E \vec y \right)^2} \Big)
	\end{equation*}
	where both summands are non-negative.
\end{lemma}

\begin{proof}
	\begin{align*}
		n^2 \Var{\vec z, \mat C}{\taunet} 
		&= n^2 \E{\vec z, \mat C}{(\taunet -\tau)^2} \\
		&= \left( \vec z' (\vec a + \vec b + 2 \mat E \vec y) (\vec a + \vec b + 2 \mat E \vec y)' \vec z \right) \tag{using the error expression, equation \ref{eq:taunet-error}} \\
		&= \E{\vec z}{\E{\mat C}{ \vec z' (\vec a + \vec b) (\vec a + \vec b)' \vec z + \underbrace{ 4 \vec z' (\vec a + \vec b) (\mat E \vec y)' \vec z}_{=0 \text{ in }\mathbb{E}_{\mat C}} + 4 \vec z' \mat E \vec y (\mat E \vec y)' \vec z } } \tag{using $\mat C \perp \vec z$}\\
		&= (\vec a + \vec b)' \Cov{\vec z} (\vec a + \vec b) + 4 \E{\vec z, \mat C}{\left( \vec z' \mat E \vec y \right)^2 }
	\end{align*}
	The second summand is clearly non-negative. For the first one note that $\Cov{\vec z}$ is positive-semidefinite.
	\end{proof}

The dependence of $\vec y$ on $\vec z$ can be written as follows:
\begin{equation}\label{eq:dependence-y-on-z}
	\vec y = \vec \mu  + \frac{1}{2} \left(\vec z \odot(\vec a-\vec b) \right)
\end{equation}
where $\odot$ represents element-wise multiplication.

This shows the challenge of the variance term in \Cref{lem:taunet-variance-form-with-expectation}: Due to the linear dependence of $\vec y$ on $\vec z$, the variance is an expectation over a \emph{quartic} term in $\vec z$. 

There are two nice properties for an expression of $\Var{\taunet}$ that we would like to achieve. 
\begin{enumerate}[label=(P\arabic*)]
\item \label{enum:nice-var-properties:positive-terms} It should only consist of positive terms, as this makes a possible algorithm design easier by avoiding the burden of cancellation. 
\item \label{enum:nice-var-properties:simple-z-dependence} Its dependence on $\vec z$ should be simpler, possibly removing the quartic dependency.
\end{enumerate}

We have explored this in two ways.
\begin{enumerate}[label=(\alph*)]
	\item By assuming natural bounds on $\vec y$, we can achieve both \ref{enum:nice-var-properties:positive-terms} and \ref{enum:nice-var-properties:simple-z-dependence}.
	\item Without making any assumptions on $\vec y$, we can achieve \ref{enum:nice-var-properties:simple-z-dependence} by  transforming the quartic dependency on $\vec z$ to a cubic dependency. However, this comes at the cost of losing \ref{enum:nice-var-properties:positive-terms}.
\end{enumerate}
Both ways (a) and (b) are described in the following sections.

\paragraph*{a) Assuming bounds on $\vec y$}~ \\
A natural assumption is a bound on the outcome $\vec y$.
\begin{assumption}[Bound on $\vec y$] \label{ass:bound-on-y} Assume
	 $\exists f \in \R^+. \forall i \in [n]$:
	\begin{equation*}
		y_i \in [\pm\sqrt{f}].
	\end{equation*} 
	This extends to $a_i, b_i \in [\pm \sqrt{f}]$, as $y_i$ can be both $a_i$ or $b_i$ with positive probability $(\Pr{z_i = 1} = \frac{1}{2})$.
\end{assumption}
The mathematical effect of this assumption is to reduce the quartic dependence on $\vec z$ to a quadratic one.

\begin{theorem}\label{thm:variance-under-bounds-assumption}
Under Assumption \ref{ass:bound-on-y}, the variance of $\taunet$ is bounded by
\begin{equation*}
	\Var{\vec z,\mat C}{\taunet} \leq \E{\vec z}{\vec{z}' \mat{M}\vec{z}} = \E{\vec z}{\norm{\mat{R}\vec{z}}_2^2} 
\end{equation*}
where $\mat{M}$ is a known, positive-semidefinite matrix and $\mat{R} := \mat{M}^\frac{1}{2}$ is its root.

\end{theorem}
	
	We first show a stochastic property of $\mat E$ that will be needed in the proof of \Cref{thm:variance-under-bounds-assumption}.
	
	\begin{lemma}\label{lem:independence-of-columns-of-E} The columns of $\mat E$ are stochastically independent. 
	
	Formally, let $\vec e_i$ denote the $i$th column of $\mat E$. Then,
	\begin{equation*}
		\forall i \neq j \in [n]: \vec e_i \perp \vec e_j.
	\end{equation*} 
	\end{lemma}
	
	\begin{proof}
		By definition of $\mat E$, we have for any $E_{i,j}$	
		\begin{align}
			E_{i,j} = \sum_{k \in [n]}  \left((\mat I + \mat A)^{-1}\right)_{i,k} (\mat I + \mat C)_{k,j} -\delta(i,j).
		\end{align}
		Analogously, for any other $E_{p,q}$, we have
		\begin{align}
			E_{p,q} = \sum_{l\in [n]}  \left((\mat I + \mat A)^{-1}\right)_{p,l} (\mat I + \mat C)_{l,q} -\delta(p,q).
		\end{align}
		If those are from different columns, meaning $j \neq q$, these two terms do not share any entry of $\mat C$. As different entries of $\mat C$ are independent by our interference model assumption, $E_{i,j} \perp E_{p,q}$ follows.	
	\end{proof}
	
	\begin{proof}[Proof of \Cref{thm:variance-under-bounds-assumption}] We first find a bound on the network part of $\Var{\taunet}$ and then turn this into the bound of \Cref{thm:variance-under-bounds-assumption}.
	
	We now rewrite the network term of $\Var{\taunet}$ as a sum of positive terms.
		
	\begin{align*}
		\E{\vec z, \mat C}{\left( \vec z' \mat E \vec y \right)^2 }
		&= \E{\vec z, \mat C}{\left(  \sum_{i \in[n]}\vec z' \vec e_i y_i \right)^2} \tag{where $\vec e_i$ is the $i$th column of $\mat E$}  \\
		&= \E{\vec z, \mat C}{ \sum_{i,j \in[n]} (\vec z' \vec e_i y_i) (\vec z' \vec e_j y_j) } \\
		&= \E{\vec z, \mat C}{ \sum_{i \in[n]} (\vec z' \vec e_i y_i)^2} + \E{\vec z, \mat C}{\sum_{i \neq j \in[n]} (\vec z' \vec e_i y_i) (\vec z' \vec e_j y_j) } \tag{linearity of $\mathbb{E}$}\\ 
		&= \E{\vec z, \mat C}{ \sum_{i \in[n]} (\vec z' \vec e_i y_i)^2} + \E{\vec z}{\E{\mat C}{\sum_{i \neq j \in[n]} (\vec z' \vec e_i y_i) (\vec z' \vec e_j y_j) }} \tag{$\vec z \perp \mat C$}\\ 
		&= \E{\vec z, \mat C}{ \sum_{i \in[n]} (\vec z' \vec e_i y_i)^2} + \sum_{i \neq j \in[n]} \E{\vec z}{\E{\mat C}{(\vec z' \vec e_i y_i)}} \E{\vec z}{\E{\mat C}{(\vec z' \vec e_j y_j) }} \tag{\Cref{lem:independence-of-columns-of-E}}\\ 
		&= \E{\vec z, \mat C}{ \sum_{i \in[n]} (\vec z' \vec e_i y_i)^2} \tag{$\E{\mat C}{\vec e_i} = \vec 0$} \\
		&= \E{\vec z}{\sum_{i\in [n]} \vec z' \Cov{\vec e_i} \vec z \: y_i^2} \tag{$\vec z \perp \mat C$}\\
	\end{align*}
	
	By applying Assumption \ref{ass:bound-on-y}, this yields
	\begin{equation}\label{eq:bound-on-network-part-in-var-taunet}
		\E{\vec z, \mat C}{\left( \vec z' \mat E \vec y \right)^2 }
		\leq f \cdot \E{\vec z}{ \sum_{i\in [n]} \vec z' \Cov{\vec e_i} \vec z }.
	\end{equation}
	
	Plugging this into $\Var{\taunet}$, we obtain 
	
	\begin{align*}
		n^2 \Var{\vec z, \mat C}{\taunet} 
		&= (\vec a + \vec b)' \Cov{\vec z} (\vec a + \vec b) + 4 \E{\vec z, \mat C}{\left( \vec z' \mat E \vec y \right)^2} \\
		&\leq  (\vec a + \vec b)' \Cov{\vec z} (\vec a + \vec b) + 4 f \cdot \E{\vec z}{ \sum_{i\in [n]} \vec z' \Cov{\vec e_i} \vec z } \tag{by equation \ref{eq:bound-on-network-part-in-var-taunet}} \\
		&= \E{z}{\vec z' \left((\vec a + \vec b)(\vec a + \vec b)' + 4 f \sum_{i\in [n]} \Cov{\vec e_i}  \right) \vec z} \\
		& \leq 4 f \cdot \E{z}{\vec z' \left( \vec 1 \vec 1' +  \sum_{i\in [n]} \Cov{\vec e_i}  \right) \vec z} \tag{by Assumption \ref{ass:bound-on-y}: $\abs{\vec a + \vec b}  \leq 2 \sqrt{f} \cdot \vec 1$}.
	\end{align*}
	
	By setting 
	\begin{equation} \label{eq:def-M}
		\mat M := \frac{4 f}{n^2} \left( \vec 1 \vec 1 '  + \sum_{i\in [n]} \Cov{\vec e_i} \right)
	\end{equation}
	
	and by realizing that all summands are positive-semidefinite, we see that their sum, $\mat M$, is also positive-semidefinite. Therefore, it is possible to take the root $\mat R:= \mat M^{\frac{1}{2}}$.
	With this choice of $\mat R$, \Cref{thm:variance-under-bounds-assumption} follows.
\end{proof}

Note that the bound of \Cref{thm:variance-under-bounds-assumption} is tight in the following sense: There exist $\vec a, \vec b$ such that it holds with equality (namely $\vec a = \vec b = \sqrt{f} \cdot \vec 1$). 

To minimize the variance of $\taunet$ in this setting, we thus need to minimize the euclidean norm
\begin{equation} \label{eq:var-bound-euclidean-norm}
	\E{\vec z}{\norm{\mat{R}\vec{z}}_2^2} \quad \text{with }\vec z \in \{\pm 1\}^n.
\end{equation}

Dan Spielman mentions in a talk \cite{spielmandaniel-DiscrepancyTheoryRandomized-2021} a generalized version of \cite{charikar-TightHardnessResults-2011}:
\begin{theorem}[\cite{spielmandaniel-DiscrepancyTheoryRandomized-2021,charikar-TightHardnessResults-2011}] \label{thm:norm-distinction}
	Given the column vectors $\vec v_1, \ldots, \vec v_n \; \in \R^d$ with $\forall i \in [n]: \; \norm{v_i}_2 \leq 1$ of a matrix $\mat V \in \R^{d,n}$, it is $\NP$-hard to distinguish
	\begin{enumerate}
		\item[(1)] $\exists z \in \{\pm 1\}^n:$ $\mat V \vec z = \vec 0$\\ from
		\item[(2)] $\forall \vec z \in \{\pm 1\}^n:$ $\norm{\mat V \vec z}_2 \geq c \sqrt{d}$
	\end{enumerate}
	for some universal (but currently unspecified) constant $c$.
\end{theorem}

This gives us some asymptotic insight into our problem. \Cref{thm:norm-distinction} implies for the case $d = n$ that it is not efficiently possible to find a $\vec z \in \{\pm 1\}^n$ such that
\begin{equation}
	 \norm{\mat R' \vec z}_2^2 \in \oh{n}
\end{equation}
for matrices $\mat R' \in \R^{n,n}$ with maximum $l^2$ column norm $1$ \emph{in general}.\footnote{Note that we chose for the dimensions $d=n$ here.} We can transform our problem (Equation \ref{eq:var-bound-euclidean-norm}) into the above form by setting 
\begin{equation}
	\mat R' := \xi^{-1} \mat R 
\end{equation}

where $ \xi := \max_{i \in [n]} \norm{\mat R_{:,i}}_2$ is the maximum column norm. 
Therefore, we cannot hope to efficiently find a better $\vec z$ for all $\mat R \in \R^{n,n}$. 

However, it has to be noted that \Cref{thm:norm-distinction} is a statement about matrices $\mat V$ \emph{in general}. We might have a very special case here where the matrices $\mat R$ have a special form for which it is indeed possible to make the distinction. We will continue this conversation in the outlook.

Do we know of an efficient algorithm that finds an asymptotically tight $\vec z \in \{\pm 1\}^n$ for any given $\mat R'$?
We do, namely the simple i.i.d.\ design.
\begin{lemma}
	The i.i.d.\ design is asymptotically tight in the above sense,  i.e., for any $\mat R' \in \R^{n,n}$ with maximum $l^2$ column norm $1$, the i.i.d.\ design gives a $\vec z \in \{\pm 1\}^n$ s.t.
	\begin{equation*}
		\E{\vec z}{\norm{\mat R' \vec z}_2^2} \in \Oh{n}
	\end{equation*}
\end{lemma}

\begin{proof}
	For any $i \in [n]$, we have
	\begin{align}
		\E{\vec z}{(\mat R' \vec z)_i^2} 
		&= \sum_{j,k \in [n]} R'_{i,j} R'_{i,k} \underbrace{\E{\vec z}{z_j z_k}}_{= \delta(j,k)}  \tag{linearity of $\mathbb{E}$} \\
		&= \sum_{j \in [n]} {R'_{i,j}}^2.
	\end{align}
	And therefore 
	\begin{align}
		\E{\vec z}{\norm{\mat R' \vec z}_2^2}
		&= \sum_{i,j \in [n]} {R'_{i,j}}^2 \\
		&= \sum_{j \in [n]} \underbrace{\sum_{i \in [n]} {R'_{i,j}}^2}_{\leq 1 \text{ by assumption on $\mat R'$}} \\
		&\in \Oh{n}.
	\end{align}
\end{proof}

Note that this only holds because we have in our case $d=n$ for the dimension of the matrix. If we had $d < n$, the i.i.d.\ design would not give an asymptotically tight bound. However, this asymptotically tight bound could be achieved by doing a random walk based on \cite{barany-CombinatorialQuestionsFinitedimensional-1981a, beck-IntegermakingTheorems-1981}. For details, see \cite{spielmandaniel-DiscrepancyTheoryRandomized-2021}.

We conclude this section with a bound on the probability of having a ``bad'' $\taunet$.

\begin{theorem}\label{thm:iid-design-bound-on-probability-of-error}

	 Under the i.i.d.\ design, we have for any $t \geq 0$
	 \begin{equation*}
	 	\Pr{\abs{ \taunet -\tau} \geq t } \leq \frac{\frac{4f}{n} \left( 1 + \frac{1}{n} \cdot \sum_{i,j,u \in [n]} \left(\left( (\mat I + \mat A)^{-1}\right)_{i,u} \right)^2 \Var{C_{u,j}}\right)}{t^2} 
	 \end{equation*}
\end{theorem}

Note that all variables in this bound are entirely determined by the experimenter's social network assumptions ($\mat A = \E{\mat C}, \Var{C_{i,j}}$).

This theorem follows by first calculating a bound on the variance under the i.i.d.\ design (\Cref{lem:i-i-d-design-var-bound}) and applying Chebyshev's inequality.
	
	\begin{lemma} \label{lem:i-i-d-design-var-bound}
		Under the i.i.d.\ design, 
		\begin{equation*}
			\Var{\taunet} \leq \frac{4f}{n} \left( 1 + \frac{1}{n} \cdot \sum_{i,j,u \in [n]} \left(\left( (\mat I + \mat A)^{-1}\right)_{i,u} \right)^2 \Var{C_{u,j}} \right)
		\end{equation*}
	\end{lemma}
	\begin{proof}
		Recall from \Cref{thm:variance-under-bounds-assumption} that we have for the network-related part of $\Var{\taunet}$
		\begin{align*}
			\Var{\vec z,\mat C}{\taunet} \leq \E{\vec z}{\vec{z}' \mat{M}\vec{z}}.
		\end{align*}
		The bound of this lemma follows by straightforward calculation and by using the assumption of this lemma, that $\vec z$ is an i.i.d. design.
		
		\begin{align}
			\E{\vec z}{\vec{z}' \mat{M}\vec{z}}
			&= \tr{\mat M} \tag{as $\vec z$ is i.i.d.} \\
			&= \tr{\frac{4 f}{n^2} \left( \vec 1 \vec 1 '  + \sum_{i\in [n]} \Cov{\vec e_i} \right)} \tag{by definition of $\mat M$, see eq. \ref{eq:def-M}} \\
			&= \frac{4 f}{n^2} \left( n +  \sum_{i\in [n]} \tr{ \Cov{\vec e_i}}\right) = (*) \tag{by linearity of $\operatorname{tr}$}
		\end{align}
		
		Recall that $\vec e_i$ is the $i$th column vector of $\mat E$, so that 
		\begin{equation}
			\tr{ \Cov{\vec e_i}} = \sum_{j \in [n]} \Var{E_{ji}}.
		\end{equation}
		Plugging this in, yields
		\begin{align} \label{eq:detail:var-calc-dep-on-var-e}
			(*) &= \frac{4 f}{n^2} \left( n +  \sum_{i,j\in [n]} \Var{E_{i,j}}\right).
		\end{align}
		
		To calculate $\Var{E_{i,j}}$, we use \Cref{lem:k-i-j-k-identity} from \Cref{app:k-i-j-k-lemma}, which gives (by setting $k=i$):
		\begin{align} \label{eq:detail:var-calc-e-var}
			\Var{E_{i,j}} = \sum_{u \in [n]} \left(\left( (\mat I + \mat A)^{-1}\right)_{i,u} \right)^2 \Var{C_{u,j}}.
		\end{align}
		
		Combining Equations \ref{eq:detail:var-calc-dep-on-var-e} and \ref{eq:detail:var-calc-e-var} yields
		\begin{equation}
			\E{\vec z}{\vec{z}' \mat{M}\vec{z}} = \frac{4 f}{n^2} \left( n +  \sum_{i,j\in [n]}  \sum_{u \in [n]} \left(\left( (\mat I + \mat A)^{-1}\right)_{i,u} \right)^2 \Var{C_{u,j}}\right)
		\end{equation}
		and the lemma follows.

	\end{proof}	

\begin{proof}[Proof of \Cref{thm:iid-design-bound-on-probability-of-error}]
	The theorem follows by plugging the bound for $\Var{\tauht}$ in Chebyshev's inequality:
	\begin{align*}
		\Pr{\abs{ \taunet -\tau} \geq t }
		&= \Pr{\abs{ \taunet -\E{\taunet}} \geq t } \tag{recall: $\E{\tauht} = \tau$} \\
		&\leq \frac{\Var{\taunet}}{t^2} \tag{Chebyshev} \\
		&\leq \frac{\frac{4f}{n} \left( 1 + \frac{1}{n} \cdot \sum_{i,j,u \in[n]}  \left(\left( (\mat I + \mat A)^{-1}\right)_{i,u} \right)^2 \Var{C_{u,j}} \right)}{t^2} \tag{\Cref{lem:i-i-d-design-var-bound}}
	\end{align*}
\end{proof}

\begin{remark}
	Assumption \ref{ass:bound-on-y} can be easily made narrower by assuming bounds \emph{per unit $i$}, i.e. $\forall i \in [n] \; \exists f_i \in \R^+. \quad y_i \in [\pm\sqrt{f_i}]$. This generalization carries simply over throughout the calculations, yielding $\mat M =  \frac{4}{n^2} \left( \vec f \vec f' + \sum_{i\in [n]} f_i \Cov{\vec e_i}\right)$ where $\vec f$ is the vector consisting of the individual bounds $f_i$.
\end{remark}

\paragraph*{b) No assumptions on $\vec y$}~ \\
If we do not make any assumptions on $\vec y$, we can work towards \ref{enum:nice-var-properties:simple-z-dependence} by transforming  the variance of $\taunet$ to have only a cubic, instead of a quartic, dependence on $\vec z$. However, we lose during this process the consisting-of-positive-terms-only property \ref{enum:nice-var-properties:positive-terms}. 
In this section, we will first establish an exact term for the variance. Based on this, we will determine the variance of $\taunet$ under an i.i.d. design.
\begin{theorem}\label{thm:var-network-part-under-dependence}
	For the network-related part of $\Var{\taunet}$, we have
	
	\begin{equation*}
		\frac{4}{n^2}\E{\vec z, \mat C}{\left(\vec z' \mat E \vec y \right)^2} = \frac{2}{n^2} \sum_{i,j,k \in [n]}{K_{i,j,k} \left( \Cov{z_i,z_k} \left(a_j^2 + b_j^2 \right) + \E{\vec z}{z_i z_j z_k} \left( a_j^2 -b_j^2\right)  \right)}
	\end{equation*}
	where the network-related constant $K$ is defined as
	\begin{equation*}
		K_{i,j,k} := \sum_{u \in [n]} \left((\mat{I} + \mat{A})^{-1}\right)_{i,u}  \:  \left((\mat{I} + \mat{A})^{-1}\right)_{k,u} \: Var[\mat{C}_{u,j}] \:.
	\end{equation*}
\end{theorem}

To prove this theorem, we first need a technical lemma:
\begin{restatable}{lemma}{kijkidentity} \label{lem:k-i-j-k-identity}
	\begin{equation*}
		\Cov{E_{i,j},E_{k,j}} = \E{\mat C}{E_{i,j} E_{k,j}} = K_{i,j,k}
	\end{equation*}
	where $ K_{i,j,k}$ is defined as in \Cref{thm:var-network-part-under-dependence}.
\end{restatable}

We prove this technical lemma in \Cref{app:k-i-j-k-lemma}.

\begin{proof}[Proof of \Cref{thm:var-network-part-under-dependence}]
	Recall from Equation \ref{eq:dependence-y-on-z} the identity for $\vec y$ that exhibits the dependence on $\vec z$:
	\begin{equation*}
	\vec y = \vec \mu  + \frac{1}{2} \left(\vec z \odot(\vec a-\vec b) \right).
	\end{equation*}
	
	We can use this identity in combination with \Cref{lem:k-i-j-k-identity} to show the theorem: 
		
	\begin{align*}
		&\quad \E{\vec z, \mat C}{\left(\vec z' \mat E \vec y\right)^2} \\
		&=\E{\vec z, \mat C}{\sum_{i,j,k,l \in [n]} z_i z_k E_{i,j} E_{k,l} y_j y_l} \\
		&=\sum_{i,j,k,l \in [n]}\E{\mat C}{ E_{i,j} E_{k,l}} \E{\vec z}{z_i z_k y_j y_l} \tag{as $\vec z \perp \mat C$} \\
		&=\sum_{i,j,k \in [n]}\Cov{\mat C}{ E_{i,j} E_{k,j}} \E{\vec z}{z_i z_k y_j^2} \tag{as $E_{i,j} \perp E_{k,l}$ for $j \neq l$ (\Cref{lem:independence-of-columns-of-E}) and $\E{\mat E} = \mat 0$}  \\
		&=\sum_{i,j,k \in [n]}\Cov{\mat C}{ E_{i,j} E_{k,j}} \E{\vec z}{z_i z_k \mu_j^2 + \frac{1}{4}z_i z_k \underbrace{z_j^2}_{=1} (a_j-b_j)^2 + z_i z_k z_j \mu_j (a_j-b_j) } \tag{using above identity for $\vec y$} \\
		&= \frac{1}{2} \sum_{i,j,k \in [n]}{K_{i,j,k} \left( \Cov{z_i,z_k} \left(a_j^2 + b_j^2 \right) + \E{\vec z}{z_i z_j z_k} \left( a_j^2 -b_j^2\right)  \right)}. \tag{by \Cref{lem:k-i-j-k-identity} and definition of $\vec \mu$}
	\end{align*}
	The theorem follows by multiplying by $\frac{4}{n^2}$.
	
\end{proof}

\begin{corollary}\label{cor:exact-bound-variance-network-part}
For the i.i.d.\ design,
\begin{equation*}
	\frac{4}{n^2}\E{\vec z, \mat C}{\left(\vec z' \mat E \vec y \right)^2} 
	= \frac{2}{n^2} \left( \sum_{i,j \in [n]} K_{i,j,i} (a_j^2 - b_j^2) -\sum_{i \in [n]} K_{i,i,i} (a_i^2 + b_i^2)  \right)
\end{equation*}
\end{corollary}

\begin{proof}
	Observe that under the i.i.d.\ design, 
	\begin{align*}
		 \Cov{z_i,z_k} &= \E{\vec z}{z_i z_k} =  \delta_{i,k} \\
		 \E{\vec z}{z_i z_j z_k}  &= -\mathbbm{1}_{i=j=k}
	\end{align*}
	due to independence of the entries of $\vec z$.
	Applying this to \Cref{thm:var-network-part-under-dependence} gives the desired result.
\end{proof}

\begin{remark}
	This exact expression can be turned into a better bound on $\Pr{\abs{ \taunet -\tau} \geq t }$ than the one given by \Cref{thm:iid-design-bound-on-probability-of-error} using the Chebyshev inequality. We do not carry this out because this bound would require knowledge of $\vec a$, $\vec b$, which is unknown to the experimenter. However, it shall be said that if the experimenter finds other bounds for $\vec a$, $\vec b$ than suggested in the previous section, they can use this exact expression (\Cref{cor:exact-bound-variance-network-part}) and apply their bounds from here on.
\end{remark}

\section{Optimizing the Worst-Case} \label{sec:worst-case-optimization-lp}

This section aims to derive a formulation similar to the formulation in Candogan et al. \cite{candogan-NearOptimalExperimentalDesign-2021}. We demonstrate thereby that under our model assumptions and estimator, it is possible to derive a similar linear program. However, as in their paper, it still suffers from major drawbacks and can only be used for small experiment sizes $n$ and under limiting assumptions on $\vec a$, $\vec b$.

One notion of a design's robustness is to have minimal variance in the worst-case potential outcomes $\vec a, \vec b$. Recall that the design $D$ is the distribution of the random assignment vector $\vec z$: $\vec z \sim D$. 

Constraining $\vec a, \vec b$ to some range, e.g. $[0,1]^n$, finding a robust design could be formulated as follows, based on \Cref{thm:var-network-part-under-dependence}:

\begin{align*}
	D_{opt} &= \argmin_{D}  \max_{\vec a,\vec b \in [0,1]^n} \Bigg( (\vec a + \vec b)' \Cov{\vec z}(\vec a + \vec b)  \\
	&+ 2 \sum_{i,j,k \in [n]}  K_{i,j,k} \left( \Cov{z_i, z_k} \left(a_j^2 + b_j^2 \right) + \E{\vec z \sim D}{z_i z_j z_k} \left(a_j^2 - b_j^2 \right) \right) \Bigg) \tag{minimum worst-case variance}\\
	\text{subject to}\\
	\E{\vec z \sim D}{\vec z} &= \vec 0 \tag{unbiasdness}
\end{align*}

However, it is not clear how to optimize this min-max expression. If we constrain the potential outcomes to be binary $\vec a, \vec b \in \{0,1\}^n$, we can formulate finding an optimal robust design as a linear program, similar to Candogan et al.:

We first introduce a technical notation that simplifies the linear program notation later. \\Let us enumerate all $2^n$ possible assignment vectors $\vec z \in \{-1,1\}^n$: $1, \ldots, 2^n$ in some order and stack all of these assignment vectors in that order on top of each other. We call this matrix of stacked vectors $\mat W \in \{\pm1\}^{2^n, n}$:
\begin{equation*}
		\mat W = \begin{pmatrix}
			-1 & \cdots & -1 & -1 & -1  \\
			-1	& \cdots & -1 & -1 & +1 \\
			-1	& \cdots & -1 & +1 & -1 \\
			-1	& \cdots & -1 & +1 & +1 \\
			\vdots &	\vdots & \ddots  & \vdots \\
			+1 & \hdots & +1 & +1 & +1 			
		\end{pmatrix}
		\quad
		\begin{matrix}
			 \text{assignment \#}1 \\
			\text{assignment \#}2\\
			\text{assignment \#}3\\
			\text{assignment \#}4\\
			\vdots \\
			 \text{ assignment \#}2^n.		
		\end{matrix}
\end{equation*}

Using this enumeration, we can describe a design $D$ as a collection of probability distributions $\{ p_u: u \in [2^n] \}$, where each $p_u$ is the probability of having \emph{that} assignment $u$. The matrix $W$ lets us write properties such as unbiasedness in an easy way.

\begin{lemma}

For $\vec a, \vec b \in \{0,1\}^n$, the following linear program gives an optimal worst-case design. Optimal worst-case design means a design for which $\tauht$ is unbiased and its variance is  minimal under the worst-case potential outcome vectors $\vec a, \vec b \in \{0,1\}^n$. 

The probability distribution $\{ p_u: u \in [2^n] \}$ describes the design, where $p_u$ is the probability of assignment u, according to our enumeration.

\begin{align*}
\displaystyle
\begin{array}{lr@{}llr}
\displaystyle{\mathop{\textup{minimize}}_{v, p_1, \ldots, p_{2^n}}}  & v \hspace{0.9em} \mbox{ } &  \\
\textup{subject to}& v \geq & \displaystyle\frac{1}{n^2}\Bigg[ \displaystyle\sum_{i,j \in [n], u \in [2^n]} p_u W_{u,i} W_{u,j} (a_i + b_i) (a_j + b_j)  &\\ 
& &\qquad + 2 \displaystyle\sum_{i,j,k \in [n]} K_{i,j,k} \Big( \displaystyle\sum_{u \in [2^n]} p_u W_{u,i} W_{u,k}(a_j^2 + b_j^2)  & \\
&  &\quad\qquad +  \displaystyle\sum_{u \in [2^n]} p_u W_{u,i} W_{u,k} W_{u,k} (a_j^2 - b_j^2) \Big) \Bigg] & \forall \vec a, \vec b \in \{0,1\}^n & (1)\\
&\displaystyle\sum_{u \in [2^n]} p_u W_{u,i} =&0&\forall i \in [n] &(2) \\
                 & \displaystyle\sum_{u \in [2^n]} p_u    =                                           & 1& &(3)\\
                 & 0 \leq  p_u \leq &1 & \forall u \in [2^n] & (4)\\
\end{array}
\end{align*}

\end{lemma}

\begin{proof}

We will first show that the above linear program gives a valid distribution, then we show that this distribution ensures that $\taunet$ is unbiased, and finally show the optimality claim.

	Constraints (3) and (4) ensure that $\{p_u: u \in [2^n] \}$ is a valid probability distribution over the sample space of all $2^n$ possible assignments.
	
Note that we can rewrite the expectation of $z_i$ 
\begin{equation}
	\E{\vec z \sim D}{z_i} = \sum_{u \in 2^n} p_u W_{u,i},
\end{equation} 

which holds by definition of $\mat W$.
Therefore, constraint (2) ensures unbiasedness, as $\taunet$ is unbiased for $\E{\vec z} = \vec 0$ (\Cref{thm:taunet-unbiased}).

It remains to show that $D$ is worst-case optimal. Note that $v$ is just a scalar, introduced for minimization. Similar to the expectation of $z_i$, we can use $\mat W$ to rewrite
\begin{align}
	\E{\vec z \sim D}{z_i,z_k} 
	&= \Cov{z_i, z_k} \tag{as $\E{\vec z \sim D}{\vec z} = \vec 0$} \\
	&= \sum_{u \in 2^n} p_u W_{u,i} W_{u,k}, \label{eq:detail:lp-prog-exp-2}\\
	\E{\vec z \sim D}{z_i,z_j,z_k} &=  \sum_{u \in 2^n} p_u W_{u,i} W_{u,j} W_{u,k}.  \label{eq:detail:lp-prog-exp-3}
\end{align}

Therefore, 
\begin{align*}
	n^2 \Var{\vec z, \mat C}{\taunet} &=  (\vec a + \vec b)' \Cov{\vec z}(\vec a + \vec b)\\ &\quad +2 \sum_{i,j,k \in [n]}  K_{i,j,k} \Bigg[ \Cov{z_i, z_k} \left(a_j^2 + b_j^2 \right) + \E{\vec z \sim D}{z_i z_j z_k} \left(a_j^2 - b_j^2 \right) \Bigg] 
	\tag{\Cref{thm:var-network-part-under-dependence}, \Cref{lem:taunet-variance-form-with-expectation}} \\
	&= \sum_{i,j \in [n], u \in [2^n]} p_u W_{u,i} W_{u,j} (a_i + b_i) (a_j + b_j)   \\
	&\quad + 2 \sum_{i,j,k \in [n]} K_{i,j,k} \Bigg[  \displaystyle\sum_{u \in [2^n]} p_u W_{u,i} W_{u,k}(a_j^2 + b_j^2) \\
	&\quad +  \sum_{u \in [2^n]} p_u W_{u,i} W_{u,k} W_{u,k} (a_j^2 - b_j^2) \Bigg] \tag{by Equations \ref{eq:detail:lp-prog-exp-2}, \ref{eq:detail:lp-prog-exp-3}}
\end{align*}
Therefore, constraint (1) is the variance term.
As the scalar $v$ is guaranteed to be greater than or equal to the variance term for all possible outcomes $\vec a, \vec b \in \{\pm 1\}^n$, $v$ is at least the worst-case variance. As we minimize $v$, it is the optimal worst-case variance.

\end{proof}

	\paragraph*{Remark on usefulness.} We emphasize this section's introductory remarks: As with Candogan et al.'s linear program, this one is only feasible for very small $n$. Note that we have not only exponentially many constraints -- which could allow for an efficient linear program if a separation oracle was found -- but also exponentially many variables $\{p_u:\; u \in [2^n]\}$. 

\section{Disconnected Graphs: Block design} \label{sec:disconnected-graph-block-design}

We have seen before that the variance of an unbiased estimator decreases naturally with $\frac{1}{n}$. The calculation (Equation \ref{eq:variance-decreases-with-high-n}), however, assumed independence between the units (SUTVA). 

It is not clear how this carries over to the non-SUTVA case,\footnote{Even though there is a $\frac{1}{n^2}$ factor in the variance term, it is not clear whether or not $\E{\vec z, \mat C}{\left( \vec z' \mat E \vec y \right)^2 } \in \Om{n^2}$} where dependence between the units is \emph{the} characteristic.

In this section, we show that if a graph has multiple connected components, the network's contribution to the variance term can be regarded as a contribution \emph{per connected component}. Therefore, we have a notion of independence at the inter-connected-components-level.

\paragraph*{Preliminaries.}
When speaking about connected components, we mean connected in the graph $G$ with adjacency matrix $A$.
Let $\mathrm{CC}(i)$ denote the index set of vertices in the connected component of vertex $i$. Let $\mathscr{S}$ be the set of connected components for a given graph. Without loss of generality, let the vertex indices be grouped by connected component, i.e. $\mathrm{CC}(1) = \{1, \ldots, k\}$, $\mathrm{CC}(k+1) = \{k+1, \ldots, l\}$, $\ldots$ for some indices $k < l \in [n]$. Therefore, $\mat A$ and $\mat C$ are block-diagonal matrices\footnote{Block-diagonality of $\mat A$ implies block-diagonality of $\mat C$ as $\E{\mat C} = \mat A$, and as the entries of $C$ are \emph{non-negative} random variables}, with each block representing a connected component.

\paragraph*{Expressing the network contribution per connected component.}
Intuitively, disconnected components should be independent terms in the network-related part of the variance. Mathematically, this stems from the block diagonality of $\mat A$. There are two ways to show this formally: We can either reason about the block diagonality of $\mat E$, or the network constant $K_{i,j,k}$. We decided for the latter, as we can build on the variance term from \Cref{thm:var-network-part-under-dependence}. 

\begin{lemma} \label{lem:K-property-disconnected}
	\begin{equation}
		K_{i,j,k} = 0 \quad \text{if} \quad \mathrm{CC}(i) \:\cap\: \mathrm{CC}(j) \:\cap\: \mathrm{CC}(k) \: = \: \varnothing
	\end{equation}
\end{lemma}

\begin{proof}
	Recall the definition of $K$:
	\begin{equation}
		K_{i,j,k} := \sum_{u \in [n]} ((\mat{I} + \mat{A})^{-1})_{i,u}  \:  ((\mat{I} + \mat{A})^{-1})_{k,u} \: Var[\mat{C}_{u,j}] \:.
	\end{equation}
	As $\mat A$ is block-diagonal, so is $\mat I+\mat A$, and thus 
	
	\begin{equation}
		(\mat{I}+\mat{A})^{-1} = \begin{pmatrix}
			(\mat{I}+\mat{A_1})^{-1} & 0 & \hdots & 0 \\
			0	& (\mat{I}+\mat{A_2})^{-1} & \hdots & 0 \\
			\vdots &	\vdots & \ddots  & \vdots \\
			0 & 0 & \hdots & (\mat{I}+\mat{A_l})^{-1}
			
		\end{pmatrix}
	\end{equation}
	where $\mat{A_1}, \ldots \mat{A_l}$ are the diagonal blocks of $\mat{A}$. 
	
	Therefore, for any $i,j \in [n]$:
	\begin{equation}
		\mathrm{CC}(i) \neq \mathrm{CC}(j) \Rightarrow ((\mat{I} + \mat{A})^{-1})_{i,j} = 0.
	\end{equation}
	
	We also have for any disconnected $i,j \in [n]$ that $\mat{A}_{i,j}=0$, and thus (by non-negativity of $\mat{C}_{i,j}$):
	\begin{equation}
		\mathrm{CC}(i) \neq \mathrm{CC}(j) \Rightarrow \Var{}{\mat{C}_{i,j}} = 0.
	\end{equation}
	
	In the expression for $K_{i,j,k}$ we therefore have 
	\begin{equation} \label{eq:detail:inner-term-of-k-i-j-k}
	\mathrm{CC}(i) \:\cap\: \mathrm{CC}(j) \:\cap\: \mathrm{CC}(k) \:\cap\: \mathrm{CC}(u) \: = \: \varnothing \; \Rightarrow \;
		((\mat{I} + \mat{A})^{-1})_{i,u}  \:  ((\mat{I} + \mat{A})^{-1})_{k,u} \: Var[\mat{C}_{u,j}] \:= 0
	\end{equation}
	because at least one of the factors must be zero, for any of the above reasons.
	
	The lemma follows by summing expression \ref{eq:detail:inner-term-of-k-i-j-k} over all $u \in [n]$.
	
\end{proof}

By \Cref{lem:K-property-disconnected}, we can express the network contribution per component by
\begin{align}
	\frac{4}{n^2} \E{\vec z, \mat C}{(\vec z' \mat E \vec y)^2} =  \frac{2}{n^2} \sum_{S \in \mathscr{S}} \sum_{j \in S} & \left( a_j^2 \sum_{i,k \in S} K_{i,j,k} \left( \Cov{z_i,z_k} + \E{z_i z_j z_k} \right) \right.  \\ 
	+ & \left.  b_j^2 \sum_{i,k \in S} K_{i,j,k} \left( \Cov{z_i,z_k} - \E{z_i z_j z_k} \right) \right)
\end{align}

and thus get for the variance of our estimator:
\begin{theorem}
	For a graph with connected components $\mathscr{S}$, the variance of the estimator $\taunet$ is
	
	\begin{align}
		 \Var{\vec z, \mat C}{\taunet} &= \frac{1}{n^2} \Bigg[ (\vec a + \vec b)' \Cov{\vec z} (\vec a+ \vec b) \\
		& + 2 \sum_{S \in \mathscr{S}} \sum_{j \in S} \left( a_j^2 \sum_{i,k \in S} K_{i,j,k} \left( \Cov{z_i,z_k} + \E{z_i z_j z_k} \right) \right.  \\ 
	& +  \left.  b_j^2 \sum_{i,k \in S} K_{i,j,k} \left( \Cov{z_i,z_k} - \E{z_i z_j z_k} \right) \right) \Bigg]
	\end{align}
\end{theorem}

\section{Outlook} \label{sec:outlook}

There are many questions worth exploring further. 

\paragraph*{Capturing ``more'' graph influence.} 
An interesting generalization of our model would be trying to capture degree-$k$ influence: In reality, a unit gets influenced from its neighbors. This unit, again, influences its neighbors -- but now with the influence it gained so far. If we consider influence from units upto the $k$-th neighbor (``degree-$k$ influence''), we would get a model like this:

\begin{equation}
	\vec y' = \sum_{i = 0}^{k}\mat C^i \vec y
\end{equation}

More realistically, spillover effects from neighbors further away are less influential, thus motivating a damping factor $\lambda < 1$. In addition, an accurate model would capture the influence from neighbors of any distance: $k \rightarrow \infty$. This gives rise to the following model:
\begin{equation}
	\vec y' = \sum_{i = 0}^{\infty}\lambda^i\mat C^i \vec y =  \left(\mat I - \lambda \mat C \right)^{-1} \vec y,
\end{equation}

which is similar to a model in the game-theoretic paper by Ballester et al.\ \cite{ballester-WhoWhoNetworks-2006a}.
While it is very nice to have an explicit representation of the matrix series, further calculations would require taking the expectation of an inverse of a matrix, making mathematical reasoning hard.

\paragraph*{An asymptotic lower bound result.}
Another interesting question comes from the complexity side: Does the hardness result from \Cref{thm:norm-distinction}, which is a statement on matrices in general, also apply to the matrices $\mat R$ of interest for our problem? This could be approached by trying to find a reduction from  \textsc{Max-2-2-Set-Splitting} -- which is known to be \NP-hard \cite{guruswami-InapproximabilityResultsSet-2004,charikar-TightHardnessResults-2011} --  to our variance bound minimization problem (Equation \ref{eq:var-bound-euclidean-norm}). If possible, \textsc{Max-2-2-Set-Splitting} could be reduced to an instance of our problem for a certain family of graphs, using a reduction similar to \cite{charikar-TightHardnessResults-2011}. It shall be noted that this statement would still give an \emph{asymptotical} lower bound, given the \emph{boundness assumptions} on $\vec y$. While experimenters are sometimes interested in small constants, such an asymptotic result would give guidance for further algorithm searches.

\chapter{Conclusion}

In this work, we covered the randomized controlled trial, one of the major scientific tools in medicine, behavioral economics and other scientific areas. We gave a formal introduction, covering terms and notions from different disciplines. Further, we introduced Harshaw et al.'s Gram-Schmidt Walk Design, built an intuition for the algorithm, and gave major theorems.

Furthermore, we developed a model in the presence of social spillover (network interference), i.e., when Rubin's Stable Unit Treatment Value Assumption (SUTVA) does not hold. For this case, we introduced an unbiased estimator and analyzed its variance. Under boundness assumptions on the potential outcomes, we reduced the task of variance minimization to the minimization of an $l^2$ norm of a matrix-vector product in expectation, where the random vector has to be in $\{\pm 1 \}^n$. \\
While the variance decreases with $\sim \frac{1}{n}$ if the SUTVA holds, we cannot make this argument in the presence of interference. However, we have shown that such an argument can be made with an increasing number of disconnected components in the underlying graph.

Social interference in randomized controlled trials is a source for many interesting questions. We motivate more research in both finding optimal designs in our model as well as in extending the model.

\newpage

\newpage

\printbibliography

\newpage
\appendix

\chapter{Appendix for the Gram-Schmidt Walk Design}

\section{Loewner Order for Submatrices}

The following lemma is needed in the proof of \Cref{lem:cov-z-bound-Q}.

\begin{lemma} \label{lem:loewner-order-for-submatrices}
	For any matrices $\mat A, \mat B \in \R^{n,n}$ with $\mat A \preceq \mat B $ we have for any $k \in [n]$:
	\begin{equation*}
		\mat A_{[1:k],[1:k]} \preceq \mat B_{[1:k],[1:k]}.
	\end{equation*}
\end{lemma}
\begin{proof} This lemma follows by simply applying the definition of the Loewner order and by setting certain vector entries to zero.

	Suppose $\mat A \preceq \mat B $. By definition of the Loewner order, we have that $\mat B - \mat A$ is positive semi-definite. Thus
	\begin{equation}
		\forall \vec x \in \R^n: \vec x' (\mat B  - \mat A) \vec x \geq 0.
	\end{equation}
	Thus, in particular for all $\vec \tilde x \in \R^n$ with fixed $\tilde x_{k+1}, \ldots , \tilde x_n = 0$ 
	\begin{equation}
		\vec{\tilde x}' (\mat B  - \mat A) \vec{\tilde x} \geq 0
	\end{equation} 
	holds. And therefore 
	\begin{equation}
		\forall \vec x \in \R^k: \vec x' (\mat B_{[1:k],[1:k]}  - \mat A_{[1:k],[1:k]}) \vec x \geq 0,
	\end{equation}
	which implies that $(\mat B_{[1:k],[1:k]}  - \mat A_{[1:k],[1:k]})$ is positive semi-definite and thus 
	\begin{equation*}
		\mat A_{[1:k],[1:k]} \preceq \mat B_{[1:k],[1:k]}.
	\end{equation*}
\end{proof}

\chapter{Appendix for the Estimation Under Network Effects}

\section{The Network-Constant $K_{i,j,k}$} \label{app:k-i-j-k-lemma}

\kijkidentity*
\begin{proof}
	Recall that by definition of $\mat E$, we have
	\begin{align}
		E_{i,j} &= \sum_{u \in [n]} \left( (\mat I + \mat A)^{-1} \right)_{i,u} \left( \mat I + \mat C \right)_{u,j}-\delta_{i,j} \\
		E_{k,j} &= \sum_{v \in [n]} \left( (\mat I + \mat A)^{-1} \right)_{k,v} \left( \mat I + \mat C \right)_{v,j}-\delta_{k,j}
	\end{align}
	
	This yields
	\begin{align*}
		\Cov{E_{i,j},E_{k,j}}
		&= \E{\mat C}{E_{i,j} E_{k,j}} \tag{as $\E{\mat E} = \mat 0$} \\
		&= \mathop{\mathbb{E}}_{\mat C}  \left[ 
		\sum_{u,v \in [n]} \left( (\mat I + \mat A)^{-1} \right)_{i,u} \left( \mat I + \mat C \right)_{u,j} \left( (\mat I + \mat A)^{-1} \right)_{k,v}  \left( \mat I + \mat C \right)_{v,j}
		\right. \\
		&\qquad - \delta_{i,j} \sum_{v \in [n]} \underbrace{\left( (\mat I + \mat A)^{-1} \right)_{k,v} \left( \mat I + \mat C \right)_{v,j}}_{\text{in }\mathop{\mathbb{E}}_{\mat C}: \;\; =\delta_{k,j}} \\ 
		&\qquad \left. - \delta_{k,j} \sum_{u \in [n]} \underbrace{\left( (\mat I + \mat A)^{-1} \right)_{i,u} \left( \mat I + \mat C \right)_{u,j}}_{\text{in }\mathop{\mathbb{E}}_{\mat C}: \;\; =\delta_{i,j}}
		+ \delta_{i,j}\delta_{k,j}  \right] \tag{by def. of $\mat E$} \\
		&=  \mathop{\mathbb{E}}_{\mat C}  \left[ 
		\sum_{u,v \in [n]} \left( (\mat I + \mat A)^{-1} \right)_{i,u} \left( \mat I + \mat C \right)_{u,j} \left( (\mat I + \mat A)^{-1} \right)_{k,v}  \left( \mat I + \mat C \right)_{v,j}
		\right] - \delta_{i,j} \delta_{k,j} \tag{by linearity of $\mathbb{E}$}  
	\end{align*}
	
	We know that $\left( \mat I + \mat C \right)_{u,j}$, $\left( \mat I + \mat C \right)_{v,j}$ are independent for $u \neq v$ due to our model assumption. For $u = v$, we can use the identity $\E{X^2} = \E{X}^2 + \Var{X}$ to rewrite the expectation in the above expression:
	\begin{align*}
		&\mathop{\mathbb{E}}_{\mat C}  \left[ 
		\sum_{u,v \in [n]} \left( (\mat I + \mat A)^{-1} \right)_{i,u} \left( \mat I + \mat C \right)_{u,j} \left( (\mat I + \mat A)^{-1} \right)_{k,v}  \left( \mat I + \mat C \right)_{v,j}
		\right] \\
		&= \sum_{u,v \in [n]} \left( (\mat I + \mat A)^{-1} \right)_{i,u}  \E{\mat C}{\left( \mat I + \mat C \right)_{u,j}} \left( (\mat I + \mat A)^{-1} \right)_{k,v}  \E{\mat C}{\left( \mat I + \mat C \right)_{v,j}} \\ 
		&\quad + \sum_{u \in[n]} \left( (\mat I + \mat A)^{-1} \right)_{i,u} \left( (\mat I + \mat A)^{-1} \right)_{k,u}  \Var{\mat C}{\left( \mat I + \mat C \right)_{u,j}}\\
		 &= \delta_{i,j} \delta_{k,j} + \sum_{u \in[n]} \left( (\mat I + \mat A)^{-1} \right)_{i,u} \left( (\mat I + \mat A)^{-1} \right)_{k,u}  \Var{\mat C}{\left( \mat I + \mat C \right)_{u,j}} \\
		 &= \delta_{i,j} \delta_{k,j} + \sum_{u \in[n]} \left( (\mat I + \mat A)^{-1} \right)_{i,u} \left( (\mat I + \mat A)^{-1} \right)_{k,u}  \Var{\mat C}{ \mat C_{u,j}}.
	\end{align*}
	
	Combining this with the above expression, we get 
	\begin{align*}
		\Cov{E_{i,j},E_{k,j}}
		&= \sum_{u \in[n]} \left( (\mat I + \mat A)^{-1} \right)_{i,u} \left( (\mat I + \mat A)^{-1} \right)_{k,u}  \Var{\mat C}{\mat C_{u,j}} \\
		&= K_{i,j,k}. \tag{by def. of $K_{i,j,k}$}
	\end{align*}
\end{proof}

\end{document}